\documentclass[twocolumn, preprint2]{aastex631}
\accepted{April 3, 2022}

\submitjournal{ApJ}
\usepackage{xspace}
\usepackage{multirow}
\usepackage{natbib}
\newcommand{\SvH}{\mbox{[S II]/H$\alpha$}\xspace}
\newcommand{\OvH}{\mbox{[O I]/H$\alpha$}\xspace}
\newcommand{\NvH}{\mbox{[N II]/H$\alpha$}\xspace}
\newcommand{\OvHb}{\mbox{[O III]/H$\beta$}\xspace}
\newcommand{\zsun}{\mbox{Z$_\odot$}\xspace}
\newcommand{\msun}{\mbox{M$_\odot$}\xspace}
\newcommand{\mstar}{\mbox{M$_*$}\xspace}

\newcommand{\mhalo}{\mbox{M$\rm_{halo}$}\xspace}
\newcommand{\mbary}{\mbox{M$\rm_{bary}$}\xspace}
\newcommand{\hb}{\mbox{H$\beta$}\xspace}
\newcommand{\ha}{\mbox{H$\alpha$}\xspace}
\newcommand{\mbh}{\mbox{M$\rm_{BH}$}\xspace}
\newcommand{\zzero}{\mbox{\textit{z}$\sim$0}\xspace}
\newcommand{\niiplot}{NII diagnostic plot\xspace}

\newcommand{\oiplot}{OI diagnostic plot\xspace}
\newcommand{\siiandoiplot}{SII and OI diagnostic plots\xspace}
\newcolumntype{C}[1]{>{\centering\let\newline\\\arraybackslash\hspace{0pt}}m{#1}}

\defcitealias{Kewley01}{Ke01}
\defcitealias{Kauffmann03}{Ka03}
\defcitealias{Kewley2006}{K06}
\defcitealias{Eckert15}{E15}
\defcitealias{Moffett15}{M15}
\defcitealias{Eckert16}{E16}
\defcitealias{Brinchmann08}{B08}

\shorttitle{Finding Low-Metallicity $z\sim0$ Dwarf AGN Candidates}
\shortauthors{Polimera et al.}

\begin{document}

\title{RESOLVE and ECO: Finding Low-Metallicity $z\sim0$ Dwarf AGN Candidates Using Optimized Emission-Line Diagnostics}

\correspondingauthor{Mugdha S. Polimera}
\email{mugpol@unc.edu}

\author[0000-0001-6162-3963]{Mugdha S. Polimera}
\affil{ Department of Physics and Astronomy, University of North Carolina at Chapel Hill, Phillips Hall, 120 East Cameron Ave., Chapel Hill, NC 27599, USA}

\author[0000-0002-3378-6551]{Sheila J. Kannappan}
\affil{ Department of Physics and Astronomy, University of North Carolina at Chapel Hill, Phillips Hall, 120 East Cameron Ave., Chapel Hill, NC 27599, USA}

\author[0000-0002-3703-0719]{Chris T. Richardson}
\affiliation{Physics Department, Elon University, 100 Campus Drive CB 2625, Elon, NC 27244, USA}

\author[0000-0003-0402-6702]{Ashley S. Bittner}
\affil{ Department of Civil, Construction and Environmental Engineering, North Carolina State University 3125 Fitts-Woolard Hall, Raleigh, NC 27606, USA}

\author{Carlynn Ferguson}
\affil{ Department of Physics and Astronomy, University of North Carolina at Chapel Hill, Phillips Hall, 120 East Cameron Ave., Chapel Hill, NC 27599, USA}

\author{Amanda J. Moffett}
\affil{Department of Physics and Astronomy, University of North Georgia, 3820 Mundy Mill Rd., Oakwood GA 30566, USA }

\author[0000-0002-1407-4700]{Kathleen D. Eckert}
\affil{ Department of Physics and Astronomy, University of North Carolina at Chapel Hill, Phillips Hall, 120 East Cameron Ave., Chapel Hill, NC 27599, USA}

\author[0000-0001-7596-8372]{Jillian M. Bellovary}
\affil{Department of Physics, Queensborough Community College, City University of New York, 222-05 56th Ave, Bayside, NY, 11364}
\affil{Department of Astrophysics, American Museum of Natural History, Central Park West at 79th Street, New York, NY 10024, USA}
\affil{Department of Physics, Graduate Center, City University of New York, New York, NY 10016, USA}

\author[0000-0002-7001-805X]{Mark A. Norris}
\affil{ Jeremiah Horrocks Institute, University of Central Lancashire, Preston, Lancashire, PR1 2HE, United Kingdom}

\begin{abstract}
Existing star-forming vs. active galactic nucleus (AGN) classification schemes using optical emission-line diagnostics mostly fail for low-metallicity and/or highly star-forming galaxies, missing AGN in typical \zzero dwarfs. To recover AGN in dwarfs with strong emission lines (SELs), we present a classification scheme optimizing the use of existing optical diagnostics. We use SDSS emission-line catalogs overlapping the volume- and mass-limited RESOLVE and ECO surveys to determine the AGN percentage in SEL dwarfs. Our photoionization grids show that the \OvHb versus \SvH diagram (SII plot) and \OvHb versus \OvH diagram (OI plot) are less metallicity sensitive and more successful in identifying dwarf AGN than the popular \OvHb versus \NvH diagnostic (NII plot or ``BPT diagram"). We identify a new category of ``star-forming AGN" (SF-AGN) classified as star-forming by the NII plot but as AGN by the SII and/or OI plots. Including SF-AGN, we find the \zzero AGN percentage in dwarfs with SELs to be $\sim$3-16\%, far exceeding most previous optical estimates ($\sim$1\%). The large range in our dwarf AGN percentage reflects differences in spectral fitting methodologies between catalogs. The highly complete nature of RESOLVE and ECO allows us to normalize strong emission-line galaxy statistics to the full galaxy population, reducing the dwarf AGN percentage to $\sim$0.6-3.0\%. The newly identified SF-AGN are mostly gas-rich dwarfs with halo mass $ < 10^{11.5} M_\odot$, where highly efficient cosmic gas accretion is expected. Almost all SF-AGN also have low metallicities (Z $\lesssim 0.4$ Z$_\odot$), demonstrating the advantage of our method.
\end{abstract}


\section{Introduction} 
\label{sec:intro}
Over the years, strong evidence has accumulated for the presence of super massive black holes (SMBHs; \mbh $\sim 10^6 - 10^9$ \msun) at the centers of almost all giant galaxies such as the Milky Way \citep{Magorrian98, Kormendy04}. About 5-10\% of these SMBHs are traditionally classified as active galactic nuclei (AGN; actively accreting gas and dust), and nearly half show possible AGN activity when including galaxies with Composite (AGN+star formation) emission and LINERs \citep[Low Ionization Nuclear Emitting Regions; e.g.,][]{Ho97, Alexander12}. At high redshift z, SMBHs in giants are thought to be formed by mergers of lower-mass `seed' black holes (BHs) present in the central regions of dwarf galaxies. At z $\sim$ 0, dwarf galaxies remain the most abundant type of galaxies, and the intermediate mass black holes (IMBHs) they host with \mbh $\sim 10^3 - 10^5$ \msun \citep[e.g.,][]{Bellovary19,Greene20} are possible relics or analogues of high-redshift seed BHs. \\ 

IMBHs have small spheres of influence, making it hard to identify them by nuclear stellar kinematic signatures unless they are extremely nearby \citep{Greene04,Greene07, Barth08, Dong12, Reines13, Moran14, Bellovary21}. Thus, identifying IMBHs by their electromagnetic signatures when accreting as AGN in dwarfs is more efficient. The first dwarf AGN ever detected was in NGC 4395 (\mbh $\sim 10^{4} - 10^{5}$ \msun), which was confirmed as a Seyfert I type AGN \citep{Filippenko89}. Since then, several dwarf AGN have been well studied with  estimated \mbh $\sim$ $10^4 - 10^6$ \msun \citep[e.g.,][]{Barth04, Valluri05,Reines11, Reines15}. IMBH  broadline signatures have also been fortuitously found in some metal-poor dwarfs \citep[e.g.,][]{Izotov08}. The record holder for the smallest central black hole is the AGN in RGG 118 with an estimated \mbh $\sim 50,000$ \msun \citep{Baldassare15}. \\

More systematic searches in the local universe are attempting to measure the dwarf AGN frequency. These searches may place a lower limit on the BH occupation fraction (percentage of galaxies with nuclear BHs), and thus help constrain BH seed models \citep{Volonteri08, Greene12}. Systematic searches can also help quantify the importance of AGN feedback for dwarf galaxy evolution \citep{Martin19}. AGN feedback has long been considered a key ingredient in the evolution of massive galaxies. In the case of dwarfs, some studies suggest that AGN feedback is important for regulating BH growth and quenching star formation \citep[e.g.,][]{Bower06, Penny18, Koudmani19}, while others suggest that it has minimal effect in star-forming \zzero dwarfs \citep[e.g.,][]{Angles-Alcazar15, Habouzit17, Latimer19}.\\

There is no clear consensus yet on the frequency of dwarf AGN; a summary of key results is provided in Section \ref{sec:discussion}. Using mid-IR colors  from WISE data,  \citet{Sartori15} identifies a $\sim$0.4\% AGN frequency in a sample of SDSS dwarf galaxies in the local Universe. Using deeper mid-IR photometry from AllWISE and stricter signal-to-noise ratio (S/N) cuts, \citet{Hainline16} find a slightly lower mid-IR AGN percentage of $\sim$0.2\%. In contrast, \citet{Kaviraj19} find that $\sim$10-30\% of slightly higher-z dwarfs ($0.1 < z < 0.3$) are AGN by using a more relaxed mid-IR color selection criteria prescribed by \citet{Satyapal14, Satyapal18}. However, \citet{Lupi20} find a very low AGN fraction of $\sim$0.4\% for a similar sample as \citet{Kaviraj19} with better multi-wavelength cross-matching between parent surveys and higher S/N restrictions for the mid-IR WISE data (for more details, see Polimera et al. 2022, in preparation, referenced as Paper II henceforth). \\ 

In the optical, \citet{Reines13} estimate a combined AGN frequency of $\sim$1\% in \zzero SDSS dwarfs by identifying AGN and Composites from the [O~III]$\lambda 5007$/H$\beta$ vs. [N~II]$\lambda 6584$/H$\alpha$ diagnostic diagram \citep*[commonly called the BPT diagram, hereafter called the NII diagnostic plot;][]{Baldwin81}, and also by identifying broad \ha line ``wings" (implying $10^5<$ \mbh  $<10^6$ \msun). The two techniques are complementary because some spectra with broad H$\alpha$ line wings still have narrow-line \NvH ratios similar to star forming galaxies, due to the low amplitudes of the broad line flux from IMBHs \citep{Reines13}. Moreover, \NvH is a proxy for metallicity \citep{ Kewley02, Pettini04, Kewley08} and the NII plot requires high \NvH at a given \OvHb to classify galaxies as AGN. Thus, the NII plot alone is not sufficient for identifying AGN in typical low-metallicity, star-forming \zzero dwarfs. \\

Using a new metallicity-insensitive HeII diagnostic \citep[\OvHb vs HeII $\lambda 4686$/H$\alpha$;][]{Shirazi12} combined with the NII plot and mid-IR selection \citep{Jarrett11, Stern12},  \citet{Sartori15} found a total AGN frequency of $\sim$1\% in a sample of \zzero SDSS dwarfs. But this percentage is still only a lower limit due to the weak nature of the HeII $\lambda 4686$ line, the authors' exclusion of Composites and LINERs from the AGN sample, and omission of broadline detection or another means of detecting strongly star-forming AGN. Some studies use a relatively new metric $d_{BPT}$, the distance of a galaxy from the star forming locus of the BPT (NII) plot, to identify AGN in strongly star-forming dwarfs \citep{Bradford18, Dickey19}. \citet{Bradford18} find that galaxies with $d\rm_{BPT} > 0.11$ dex have some level of AGN activity, implying a dwarf AGN frequency of $\sim$3\% in a sample of isolated \zzero SDSS dwarfs that have strong emission lines and 21cm HI data. This method still does not address the metallicity bias of the NII plot.  \\

Multiple X-ray studies have estimated dwarf AGN frequencies: $\sim$0.1\% at $z < 0.3$ \citep{Schramm13}, $\sim$2\% at z $< 0.055$ \citep{Lemons15}, $\sim$0.6-3\% at $0.2<z<0.8$ \citep{Pardo16}, and $\sim$1\% at z $< 0.25$ \cite{Birchall20}. However, these studies are incomplete due to inability to detect low X-ray luminosity sources and/or inability to separate dwarf AGN from X-ray binaries (XRBs). An interesting point is that not all X-ray detected dwarf AGN would be classified as AGN by the NII plot, and conversely X-ray counterparts of optically identified dwarf AGN are rare \citep{Baldassare17}. \\ 

At the other end of the electromagnetic spectrum, \citet{Reines19} find compact radio sources corresponding to active BHs in $\sim$12\% of their 111 dwarf galaxy subsample from the SDSS NSA catalog. These radio-AGN hosts have systematically higher \OvH relative to the parent dwarf sample, and they are mostly classified as Seyferts by the \OvHb vs. \OvH plot \cite[hereafter OI plot;][]{Veilleux87} despite being classified as star-forming by the NII plot. This trend is seen with greater clarity in a follow-up study of these radio dwarf AGN by \citet{Molina21}. This result foreshadows the new AGN detection technique we present in this work. \\

The methods described above are generally biased against finding AGN in typical \zzero dwarfs in some way. Local dwarf galaxies can be defined as having baryonic mass (stellar mass + cold gas mass) \mbary $< 10^{9.9}$ \msun, equivalent to \mstar $\lesssim 10^{9.3-9.7}$ \msun \citep[the gas-richness threshold mass;][]{Kannappan13}; in this paper, we adopt \mstar $< 10^{9.5}$\msun to define dwarfs. Non-satellite dwarfs typically have high gas content \citep{Kannappan04,Kannappan13}, high star formation \citep{Geha12, Kannappan13}, and low metallicity \citep{Tremonti04, Mannucci10}. The highly star forming nature of typical dwarfs can easily dilute the spectral contribution from the lower-mass BHs that they host. Dwarf AGN hosts can have extremely blue colors on account of their strong star formation and can be misclassified as star-forming by mid-IR color selection techniques \citep{Sartori15, Hainline16}. Similarly, high-mass XRB emission -- common in highly star forming galaxies -- can hinder unambiguous detection of X-ray emission from IMBH activity  \citep[e.g.][]{Baldassare17}. Additionally, since dwarfs are metal-poor, they have low \NvH, thwarting classification using the BPT plot.  \\

To fill the need for improved detection of AGN in typical dwarfs, we propose an optimized classification scheme that uniquely classifies galaxies mainly using the metallicity-insensitive OI plot along with the metallicity-sensitive NII plot. We present updated photoionization models (see Section \ref{sec:models}) using Cloudy \citep{Ferland17} and BPASS stellar population models \citep{Stanway18} to show that the metallicity-insensitive OI plot can identify a theoretical dwarf AGN whose spectrum has up to $\sim$90\% contribution from star formation and is classified as star-forming by the NII plot. We call such galaxies `SF-AGN'. We also examine the utility of the metallicity-insensitive \OvHb vs. \SvH diagram \citep[hereafter SII plot; ][]{Veilleux87}, and find that it has mostly redundant information and is less sensitive to SF dilution compared to the OI plot. Our combination of the NII, SII, and OI plots provides an optimal method of identifying an often overlooked population of AGN in typical \zzero dwarfs.\\

We apply our new classification scheme to SDSS spectra for the mass- and volume-limited and highly complete RESOLVE survey and the less complete, but much larger, ECO survey described in Section \ref{sec:data}. Both surveys span a wide range of environments and are dominated by dwarf galaxies in low-density environments down to a baryonic mass limit of $\sim$10$^{9.2}$ \msun. About $67\%$ of RESOLVE and ECO galaxies are dwarfs. With their volume-limited design, RESOLVE and ECO present an opportunity to measure the AGN frequency in strong emission-line \zzero dwarf galaxies. We describe our optimized galaxy classification scheme in Section \ref{sec:classify}, and we argue that it identifies a new category of AGN called SF-AGN in Section \ref{sec:identify}. We show that this new SF-AGN category is a hidden population of candidate AGN in metal-poor, gas-rich, star-forming dwarfs in Section \ref{sec:physical}. Our results show that the dwarf AGN frequency among strong emission line galaxies may be much higher than previously estimated: ${\sim}3-16\%$ compared to ${<}1\%$ from previous optical searches \cite[e.g.,][]{Sartori15, Reines15}. We compare our work to previous key results in Section \ref{sec:discussion}. Finally, we summarize our conclusions and future work in Section \ref{sec:conclusion}. Throughout this paper, our analysis assumes a Hubble constant of $H_0=70$ $km s^{-1} /Mpc$. \\

\section{Data and Observations}\label{sec:data}
\subsection{The RESOLVE and ECO surveys}\label{subsec:surveydata}
The RESOLVE survey \citep[REsolved Spectroscopy Of a Local VolumE survey;][]{Kannappan08} is volume-limited in two equatorial footprints covering $>$50,000 Mpc$^3$ within a redshift range of 0.015 $< z <$ 0.023 (4500 $< cz <$  7000 $km s^{-1}$). RESOLVE-B covers a volume of $\sim$13,700 Mpc$^3$ and is highly complete down to $\rm M_r = -17.0$ as it has added redshift coverage from SDSS Stripe 82 and other sources \citep[][henceforth E15]{Eckert15}. \\ 

The ECO catalog \citep[Environmental COntext catalog;][henceforth M15]{Moffett15} is an archival, volume-limited data set designed to complement RESOLVE by performing a similar census in a volume that is an order of magnitude larger ($>$$400,000$ Mpc$^3$) in the northern spring sky, overlapping RESOLVE-A and spanning a redshift range of 0.01 $< z <$ 0.023 (3000 $< cz < $ 7000 $km s^{-1}$). \\

We define a baryonic mass-limited sample with \mbary $> 10^{9.2}$ \msun for both surveys. There are 1202 RESOLVE galaxies and 7767 ECO galaxies that satisfy this criterion. RESOLVE-B is a highly complete sample due to its overlap with Stripe 82 with added redshift coverage. ECO, however, is less complete than RESOLVE-B since it does not have additional redshift coverage and its redshift limits exclude some galaxies in massive groups/clusters with large peculiar velocities \citep[][henceforth E16]{Eckert16}. We use the highly complete RESOLVE-B sample to calculate the baryonic mass incompleteness for ECO in the same manner as \citetalias{Eckert16}, but with a baryonic mass floor of  $ 10^{9.2}$ \msun. The incompleteness for ECO is calculated as the ratio of the number of RESOLVE-B galaxies above the ECO mass floor but below the ECO luminosity floor ($\rm M_r \sim -17.33$) to the total number of RESOLVE-B galaxies above the ECO mass floor. From this, we estimate the baryonic mass completeness for ECO to be $\sim$97\%. \\

The data products of the RESOLVE and ECO surveys are mostly homogeneous with the biggest differences being the data quality and survey completeness. RESOLVE-B overlaps Stripe 82 and has deep optical data from SDSS and Medium Imaging Survey (MIS)-depth UV data from GALEX, whereas ECO (including RESOLVE-A) has only shallow optical SDSS data and mixed All Sky Imaging Survey (AIS)- and MIS-depth UV coverage, with MIS-depth for less than half of the survey footprint (including RESOLVE-A). Both RESOLVE and ECO have photometric magnitudes estimated by reprocessing existing photometry from the UV to the near-IR as described in \citetalias{Eckert15} and \citetalias{Moffett15}. The improvements to the photometric reprocessing that are applied to both surveys are described in \citetalias{Eckert15} and \citetalias{Eckert16}. Stellar masses and colors are estimated using a  Bayesian spectral energy distribution (SED) fitting code (\citealt{Kannappan13}; \citetalias{Eckert15}; \citetalias{Eckert16}). \\

For both surveys, baryonic mass, i.e., stellar mass + neutral atomic gas mass, is also estimated. For RESOLVE-B, the gas data from the 21cm HI line come from the ALFALFA survey and from new observations with the Green Bank and Arecibo telescopes \citep{Stark16}. Nearly all galaxies in the RESOLVE survey have high quality HI data, i.e., either detections or strong upper limits of $<$5-10\% of the stellar mass. ECO galaxies within the region outside RESOLVE and overlapping the ALFALFA40 public catalog have flux-limited HI data (\citetalias{Eckert16}). For 21cm observations that are missing, have weak upper limits, or cannot be deconfused,  M$\rm_{HI}$ is estimated by the photometric gas fraction method (\citetalias{Eckert15}), which uses relationships between color, axial ratio, and gas-to-stellar mass ratio. The total gas mass is estimated as 1.4M$\rm_{HI}$ to account for the Helium contribution. \\

To estimate the halo masses of galaxy groups, RESOLVE and ECO galaxies are associated to groups using the friends-of-friends group-finding algorithm described in \citetalias{Moffett15} and \citetalias{Eckert16}. Once the galaxy groups are determined, halo masses are estimated by using halo abundance matching as detailed in \citet{Eckert17}. \\

To measure the star formation activity of our sample, we use a metric called fractional stellar mass growth rate (FSMGR; the ratio of newly formed stellar mass to preexisting stellar mass divided by the timescale separating new vs. preexisting mass) instead of a specific star formation rate (sSFR), since the latter asymptotes at high SFRs \citep[see Figure 9 in][]{Kannappan13}. The long-term FSMGRs for both surveys are estimated from the same stellar population modelling code that we use to determine the stellar mass (see \citealt{Kannappan13} for more details), and is defined as: 
\begin{equation}
    \rm FSMGR_{LT} = \frac{mass_{formedinlastGyr}}{1\: Gyr \times mass_{preexisting}} 
\end{equation}

To estimate star formation rates (SFRs) for RESOLVE and ECO galaxies, we use custom reprocessed WISE mid-IR and GALEX UV photometry (Paper II and \citetalias{Eckert15}, respectively). Extinction corrections for the UV photometry are estimated from SED fitting with optical (SDSS) + UV (GALEX) + near-IR (2MASS and UKIDSS) data (\citetalias{Eckert15}). As GALEX FUV imaging is incomplete for our surveys, our default UV-based SFRs are based on NUV calibrations from \citet{Wilkins12}. Our mid-IR SFRs are based on WISE calibrations from \citet{Jarrett13}. We use the prescription from \citet{Jarrett13} to combine the non-dusty (UV) and dusty (IR) SFRs to infer total SFR (adopting $\eta = 0.17$ and $\gamma = 1$; \citealt{Buat11}). To compute SFR$\rm_{IR}$ we require galaxies to have S/N $> 5$ in the W1, W2, and W3 WISE bands, and we either omit W4 or use the data only if S/N $> 5$ in that band. For galaxies with inadequate WISE S/N , SFR$\rm_{IR}$ is not computed and the total SFR is simply SFR$\rm_{UV}$.  As expected, we find that SFR$\rm_{UV}$ is typically dominant, except for very dusty (giant) galaxies where SFR$\rm_{IR}$ contributes significantly. For this work, we use these SFRs to compute a short-term FSMGR on a timescale of 100 Myr as:
\begin{equation}
    \rm FSMGR_{ST} = \rm \frac{100\: Myr \times SFR}{[M_* - (100\: Myr \times SFR)] \times 0.1 Gyr}.
\end{equation}
We have validated SFRs for RESOLVE and ECO by comparison to SFRs from \citet{Salim16}, who also performed SED fitting with GALEX and SDSS data (see Paper II for details). RESOLVE/ECO SFRs are on average $\sim$0.2-0.3 dex higher than those from \citet{Salim16}, consistent with expectations given the improved flux recovery by the reprocessed photometry for these surveys (\citealt{Kannappan13}; \citetalias{Eckert15}). We tabulate the SFRs for all of RESOLVE and ECO in Paper II, which also examines mid-IR AGN detection. \\

Throughout this paper, we use the RESOLVE and ECO surveys in a non-duplicating way, i.e., without double-counting the galaxies common to both surveys. \\

\subsection{Emission Line Measurements and Dereddening for Strong Emission Line Sample} \label{subsec:fluxes}
\begin{figure}
    \centering
    \includegraphics[width = \linewidth]{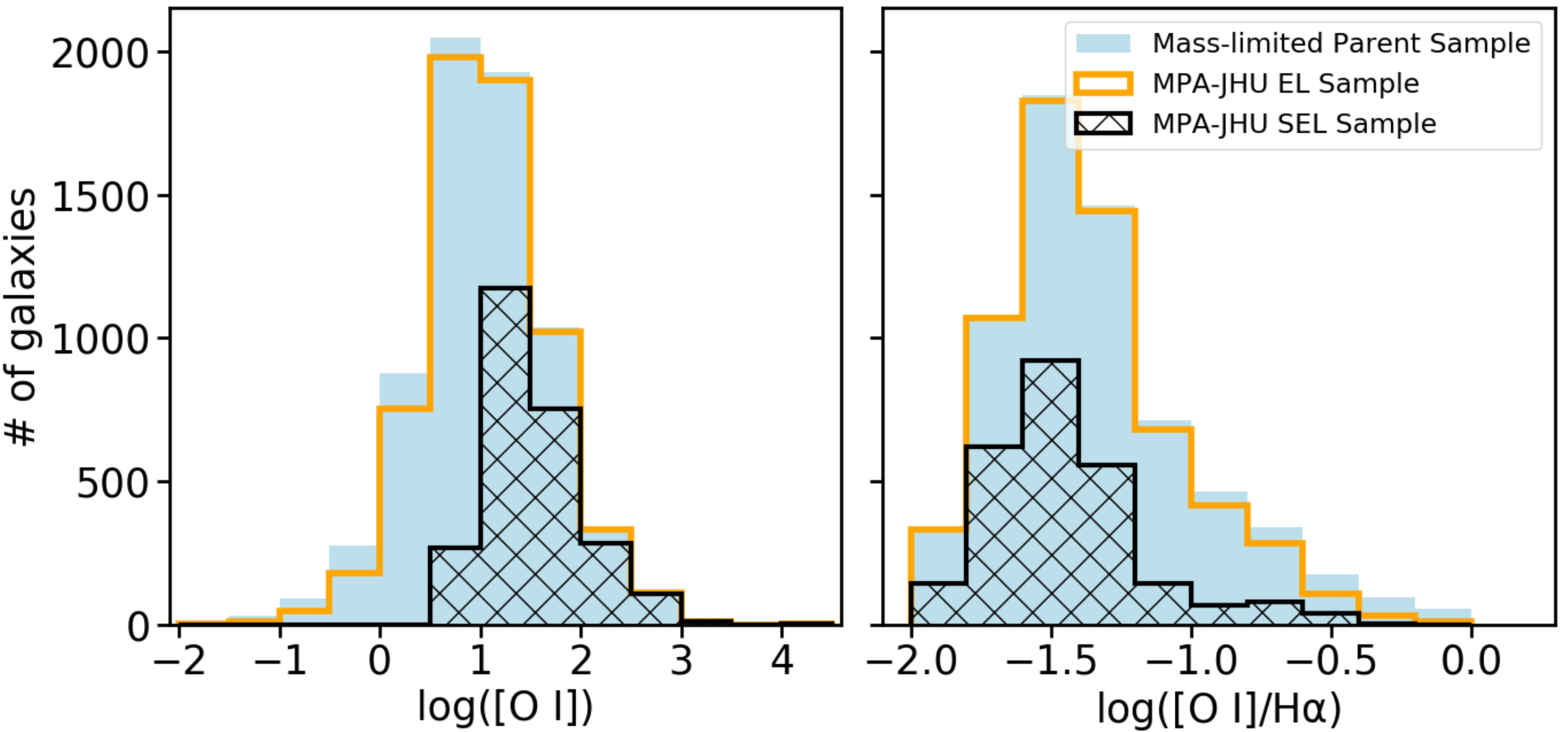}
    \caption{Distributions of [O I] line fluxes and \OvH ratios of the mass-limited parent, MPA-JHU emission line (EL; \ha S/N $>$ 5), and MPA-JHU strong emission line (SEL; S/N $> 5$ for all SELs) samples in RESOLVE and ECO combined. The high S/N$>5$ cut for the relatively weak [O I] line restricts our parent sample greatly but is required to use the emission line effectively.} \label{fig:eco+resolvesoi_hist}
\end{figure}

\begin{figure*}
    \centering
    \includegraphics[width = \linewidth]{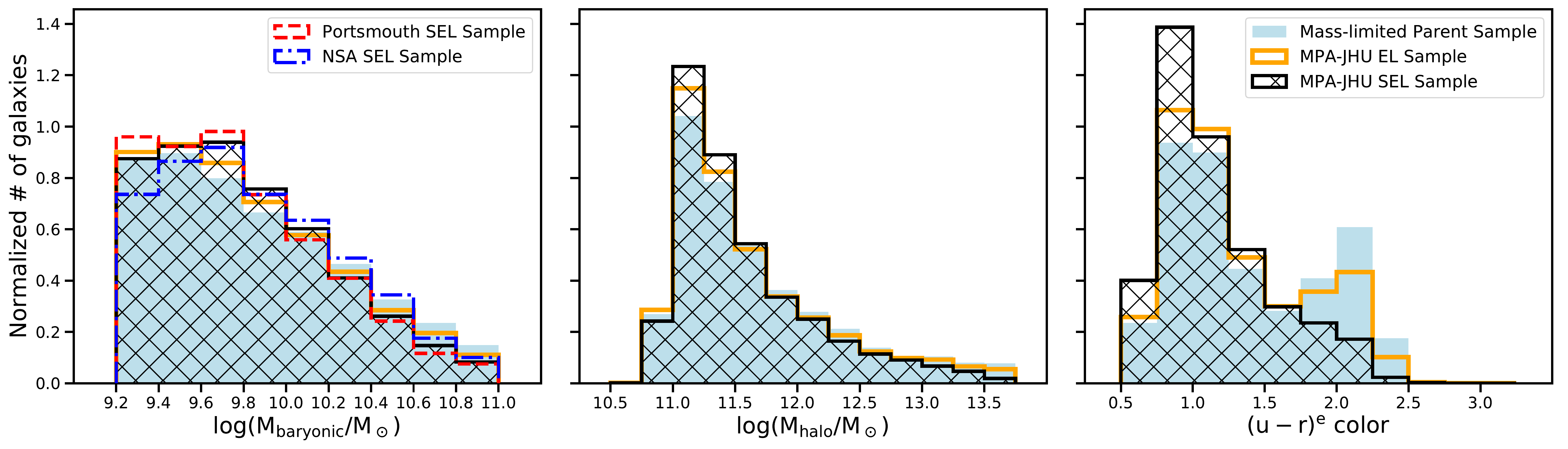}
    \caption{\textit{Left:} Baryonic mass distributions for the mass-limited parent survey and MPA-JHU EL and SEL
    samples in RESOLVE and ECO combined (light blue filled, orange open, and black cross-hatched histograms). The distributions for the Portsmouth and NSA SEL samples are shown in red and dark blue. The sharp drop off on the low mass end is the selection limit applied at \mbary $= 10^{9.2}$ \msun. The EL and SEL distributions span the mass range of the parent sample. The SEL distribution has an overdensity of relatively high-mass dwarfs (\mbary $< 10^{9.9}$ \msun) and underdensity of giants as expected.  \textit{Center:} Group halo mass distributions for the mass-limited parent survey and MPA-JHU EL and SEL samples. The EL and SEL distributions span the mass range of the parent sample. The SEL distribution has an overdensity of low-mass halos (M$\rm_{halo}\sim 10^{11.0 - 11.5}$ \msun) as expected since such halos host isolated dwarfs \citep{Eckert16}. \textit{Right:} De-extincted (u-r) color distributions for the mass-limited parent survey and MPA-JHU EL and SEL samples. The SEL sample has almost no red sequence galaxies as expected.} \label{fig:resolvesample_hist}
\end{figure*}

We are currently in the process of extracting flux measurements from new RESOLVE 2D and 3D spectroscopy (in prep). In the meantime, we have obtained results for this study using spatially unresolved flux measurements from the SDSS. We have cross-matched galaxies from the mass-limited RESOLVE and ECO surveys with three different SDSS-derived emission line catalogs. The cross-match statistics of the 1202 RESOLVE galaxies are: 1082 in the MPA-JHU catalog \citep{Tremonti04}, 1082 in the Portsmouth catalog \citep{Thomas13}, and 1075 in the NSA catalog \citep{Blanton11}. The cross-match statistics of the 7767 ECO galaxies are: 7221 in MPA-JHU catalog, 7221 in Portsmouth catalog, and 7316 in NSA catalog. The combined ECO and RESOLVE mass-limited data set (not double-counting overlap) has 8160 galaxies, and these galaxies have 7557/7557/7657 cross-matches in the MPA-JHU/Portsmouth/NSA catalogs.\\

We filter the SDSS data based on the presence of the following strong emission lines (SELs): H$\beta$, [OIII] $\lambda$5007, [OI] $\lambda$6300, [NII] $\lambda$6548, H$\alpha$, [NII] $\lambda$6584, [SII] $\lambda$6717 and [SII] $\lambda$6731. To exclude spurious measurements, we require all SELs to have a `reliable' flag in the SDSS spectroscopic catalogs and to have positive finite fluxes and errors, both of which are less than $10^5$. The subsamples that pass these cuts in RESOLVE have 901/834/760 galaxies for the MPA-JHU/Portsmouth/NSA catalogs, and in ECO have 6005/5460/5205 galaxies for the MPA-JHU/Portsmouth/NSA catalogs. To reliably use the NII, SII, and OI plots, we also apply a signal-to-noise ratio requirement of S/N $> 5$ on all the aforementioned SELs; the resulting sample is mainly limited by the S/N restriction for the weak [OI] line. These criteria reduce our mass-limited RESOLVE sample (of 1202 galaxies) to 382/202/209 SEL galaxies from the MPA-JHU/Portsmouth/NSA catalogs. Similarly, the mass-limited ECO sample (of 7767 galaxies) is reduced to 2507/1161/1363 SEL galaxies from the MPA-JHU/Portsmouth/NSA catalogs. The combined ECO and RESOLVE SEL samples, not double-counting the overlap, contain 2605/1207/1411 galaxies from the MPA-JHU/Portsmouth/NSA catalogs. Among these SEL galaxies $\sim$60\% are dwarfs in all catalogs. Figure \ref{fig:eco+resolvesoi_hist} shows that our strict S/N cuts, especially on the weak [OI] line, greatly reduce our parent survey to $\sim$16--35\% of its original size. However, a S/N $>$ 5 cut is critical to use the weak [OI] line effectively. Interestingly, the S/N cut on [OI] flux selects relatively more galaxies with \textit{low} [OI]/H$\alpha$ \textit{ratio}, which selects \textit{against} AGN in the \OvHb vs \OvH diagnostic plot, (contrary to naive intuition given the rising correlation between AGN luminosity and [O~I] line flux taken alone). In Section \ref{sec:models}, we will justify our choice to use the [O~I] line, and in Section \ref{sec:discussion}, we will show that despite the apparent selection bias against AGN implied by our S/N cuts, our new method still yields a higher AGN percentage for dwarfs than most other methods.\\

Figure \ref{fig:resolvesample_hist} shows the distributions of baryonic mass, halo mass and (u-r) color for the mass-limited parent sample, the emission line sample (EL; only \ha S/N $> 5$), and the SEL sample in ECO. The EL and SEL samples have different baryonic mass, halo mass, and color distributions than each other and than the parent survey (at high confidence $>$4$\sigma$ based on K-S tests for both samples). Nonetheless, these distributions span the full range of baryonic mass and halo mass, showing more bias in color. The differences we see are in line with expectations -- requiring the presence of H$\alpha$ emission biases the EL sample to have more dwarfs than giants and more blue galaxies than red galaxies, and further requiring the presence of all SELs biases the sample towards having more relatively high-mass dwarfs (\mbary $\sim 10^{9.6 - 9.9}$ \msun) and bluer colors. Since dwarfs are generally gas-rich and actively star-forming, and giants are generally more gas-poor and not as actively star-forming, this bias is expected by design. The RESOLVE samples also follow the same trends. Nonetheless, since the EL and SEL samples span the entire range of baryonic and halo masses in the parent surveys, we expect our statistics to offer broad insight into the underlying \zzero galaxy population.  \\

Flux measurements in the MPA-JHU and NSA catalogs are corrected for Milky Way foreground extinction. 
Fluxes from the Portsmouth catalog need additional corrections. We follow the steps outlined in \citet{Thomas13} and correct for (1) Milky Way foreground extinction, and (2) per-plate r-band flux rescaling, taking both correction factors from the MPA-JHU catalog. \\

None of the three SDSS catalogs account for galaxy internal extinction corrections, so we perform emission line dereddening using Balmer decrements to estimate internal extinction. We determine each galaxy's color excess, E(B-V), from the ratio of the galaxy's foreground extinction-corrected H$\alpha$/H$\beta$ flux ratio to the intrinsic H$\alpha$/H$\beta$ ratio of 2.86\footnote{We have also tested our classifications using \ha/\hb = 3.1, the typical Balmer decrement in an AGN ionization field \citep{Ferland86}. In this test, only 2 galaxies were assigned different classifications (one went from SF to Composite, and one from SF to SF-AGN; see Section \ref{subsec:newscheme} for category definitions). Overall, the results did not change within the error bars of the statistics quoted in Section \ref{subsec:agnstats} since the line ratios by design have very close wavelengths and are typically not affected much by extinction. In order to use different Balmer extinction values for SF and AGN galaxies, we would need an iterative process to determine the appropriate Balmer decrement for each galaxy. Since SF galaxies are the majority of the sample and the results do not change with the usage of either Balmer decrement value, we choose to use the value 2.86 for all galaxies.} \citep{Dominguez13}. For galaxies with M$_* < 10^9$ \msun, we use the SMC extinction curve given by \citet{Gordon03} with a slight modification: since their polynomial does not fit the optical data well, we fit a line to their data for wavelengths greater than 3030\AA, and we redefine the extinction curve for optical wavelengths as:
\begin{equation}
    \rm \frac{E(x - V)}{E(B-V)} = 1.91 x^2 -0.80 x - 2.75,
\end{equation}
where x $= 1/\lambda$ $\rm \mu m$. For galaxies with \mstar $ > 10^{10}$ \msun, we use an unmodified \citet{Odonnell94} Milky Way extinction curve. For galaxies with intermediate masses, we use a smoothly varying linear combination of the SMC and Milky Way extinction curves.   \\

With these corrections, fluxes from the MPA-JHU and Portsmouth catalogs are comparable, even though the Portsmouth catalog uses only solar metallicity stellar population models to fit the continuum, while the MPA-JHU catalog includes low-metallicity models (see Figure 2 in \citealt{Thomas13}). However, the flux ratios used in the NII, SII, and OI plots are on average $\sim$10\% higher from the NSA catalog than from the MPA-JHU and Portsmouth catalogs. This difference mainly arises from an additional flux calibration in the NSA catalog to fix small-scale calibration residuals \citep{Yan11}. The NSA catalog seems to have superior flux calibration, but the MPA-JHU catalog has almost twice as many galaxies, and it uses low-metallicity continuum models better suited to dwarfs (see Section \ref{subsec:catalogdiff} for more details). \textit{Since all three catalogs have matches for roughly the same fraction of RESOLVE and ECO galaxies ($\sim$90\%), we believe that the differences in SEL subsample sizes are primarily driven by differences in spectral modeling methodologies that manifest as sample selection effects when S/N cuts are imposed} (see Section \ref{subsec:catalogdiff}). As we do not find a single catalog that is \textit{clearly} better, we report statistics from all three catalogs. We use the MPA-JHU cross-matched sample for plots since it has the most RESOLVE/ECO galaxies of the three cross-matched samples.\\

The high degree of completeness and volume- and mass-limited  survey definitions  of  RESOLVE and ECO allow us to normalize statistics for the SEL samples to the full parent survey, yielding ``completeness-corrected” statistics useful for comparisons to theory \citep[e.g.,][] {Haidar22}. RESOLVE and ECO also allow us to study the properties of AGN hosts using extensive supporting data (especially gas content and environment), providing added value to SDSS emission line fluxes.   \\

\subsection{Photoionization Models} \label{sec:models}
\begin{figure*}[]
    \centering
    \includegraphics[width=\linewidth]{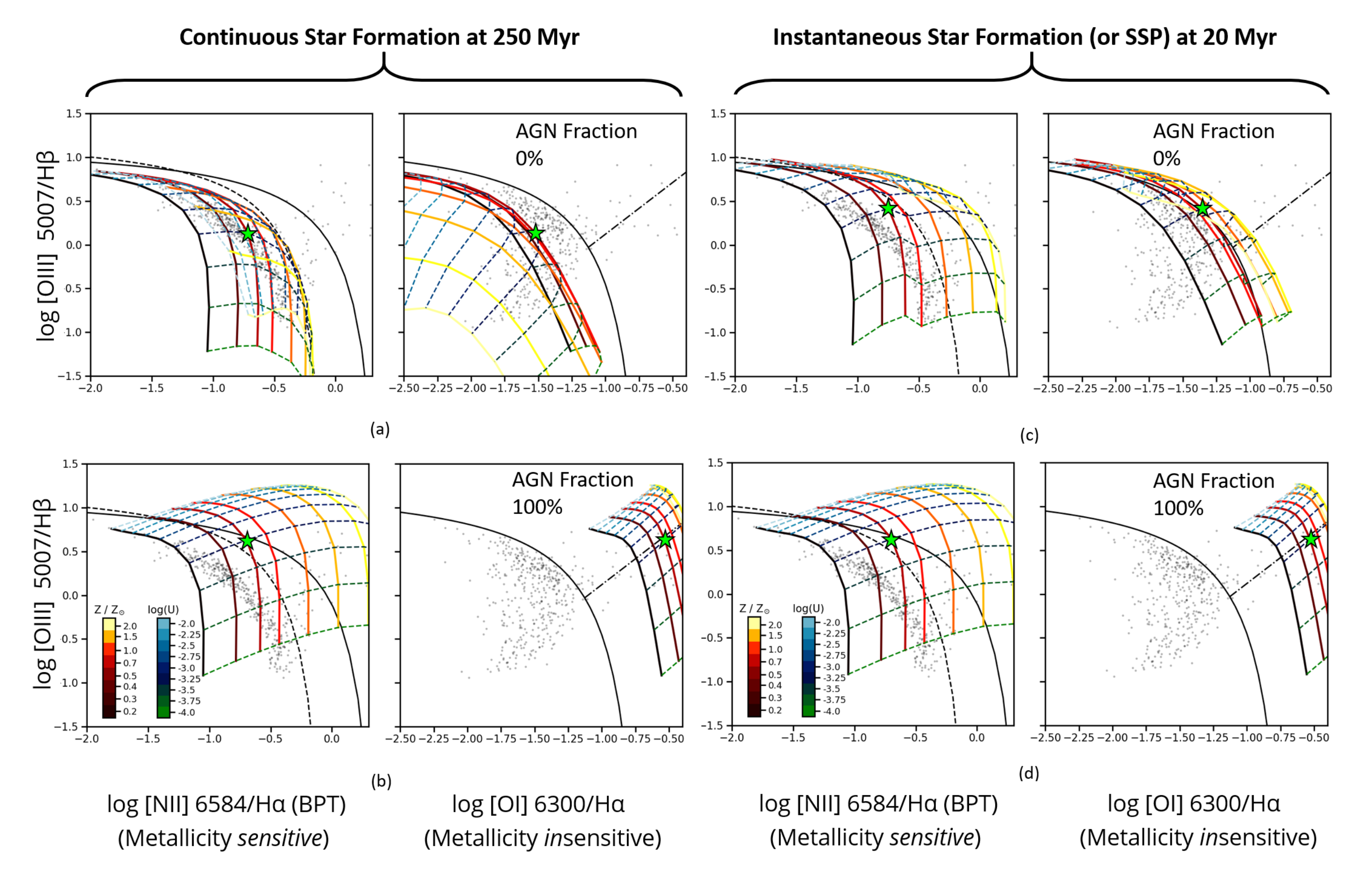}
    \caption{Photoionization grids with CSF and SSP histories at 0\% and 100\% AGN mixing fractions plotted over the emission line diagnostic plots. Lines of constant Z are drawn in the brown to yellow color-scale at [0.2, 0.3, 0.4, 0.5, 0.7, 1.0, 1.5, 2.0] Z$_\odot$ (Note: We plot a limited set of Z values from our full grid for clarity in the plots). Lines of constant log(U) are drawn in the green to blue color-scale at [$-$4.0, $-$3.75, $-$3.50, $-$3.25, $-$3.00, $-$2.75, $-$2.50, $-$2.25, $-$2.00, $-$1.75] dex. Gray dots represent RESOLVE SEL galaxies. The lime-green star represents a fiducial model dwarf with Z = 0.4 \zsun and log(U) = $-$3.25. Solid black lines represent the theoretical maximum starburst lines as given by \citet[][Ke01]{Kewley01}. Dashed lines indicate the lower edge of the `Composite' galaxy region in the NII plot as given by \citet[][Ka03]{Kauffmann03}. For grids with AGN fraction 100\% all grid points lie above the \citetalias{Kewley01} demarcation line only in the OI plot regardless of SFH. Thus, the OI plot seems better than the NII plot at identifying AGN in low metallicity, star-forming galaxies.}
    \label{fig:csfandsspgrid}
\end{figure*}
To aid in interpretation of the optical emission line diagnostic plots and estimation of gas-phase metallicities, we have computed new photoionization model grids with Cloudy \citep{Ferland17} as presented in \citet{Richardson22}. These models use the BPASS stellar population synthesis code \citep{Stanway18} to account for binary stellar populations and to take advantage of BPASS's flexibility with respect to metallicity. \\

Our grids span metallicities (Z) of [0.05, 0.1, 015, 0.2, 0.3, 0.4, 0.5, 0.7, 1.0, 1.5, 2.0] Z$_\odot$, and ionization parameter log(U) values in the range [-4.00, -1.75] in increments of 0.25 dex. We selected the Z and U values based on the availability of models in BPASS and the range of values we can potentially find in local surveys like RESOLVE and ECO. Our models include AGN mixing fraction values of [0, 4, 8, 16, 32, 50, 64, 100] percent, assuming coincident mixing with an open geometry (see \citealt{Richardson22} for more details). An AGN fraction of $8\%$ means that $8\%$ of the incident radiation field is due to the AGN, and 92\% is due to SF. To simulate IMBHs in \zzero dwarfs, we use the \mbh = $10^5$ \msun AGN SED given by the QSOSED model \citep{Kubota18}. We choose this AGN model as it is more physically realistic than other popular AGN SEDs, especially for modelling IMBHs in dwarfs (see \citealt{Richardson22} for a detailed discussion; also see \citealt{Panda19}). \\

We create two grids with different choices of star formation histories (SFHs) -- one with a continuous star formation history (CSF) viewed at 250 Myr, and one with an instantaneous SFH or simple stellar population (SSP) viewed at 20 Myr. In Figure \ref{fig:csfandsspgrid}, we show the two most important emission line diagnostic plots, the NII and OI plots, with the full photoionization grids for CSF and SSP SFHs. Grid points above the \citet[][henceforth Ke01]{Kewley01} demarcation line (solid black line) are traditionally classified as AGN. As will be discussed in Section \ref{subsec:need}, the green star (a fiducial model dwarf representative of the low metallicities of RESOLVE and ECO SEL galaxies) falls below this \citetalias{Kewley01} line even for a 100\% AGN model. In panel (a), we can see that our grids are offset $\sim$0.2-0.3 dex lower than the \citetalias{Kewley01} demarcation line, which was derived from a PEGASE-based CSF model (details in \citetalias{Kewley01}). We have tested our grids with different settings and concluded that the BPASS and PEGASE SEDs are similar, with PEGASE producing a slightly harder continuum at energies $> 10.4$keV, which may result in a portion of the offset \citep{Levesque10}. We attribute the rest of the offset to differences in abundances and depletion factors between our model and that of \citetalias{Kewley01} (see \citealt{Richardson22}).  \\

A point of concern is the fact that the maximum starburst lines of our grids do not follow the loci of data points. This difference can be attributed to the fact that real dwarfs have complicated SFHs. The choice of SFH for the model affects the predicted line ratios. For the CSF grid with 0\% AGN (Figure \ref{fig:csfandsspgrid}a), all grid points fall below the \citetalias{Kewley01} maximum theoretical starburst line in both plots, as expected, but the grid falls below the real data in the OI plot. In contrast, the SSP model with 0\% AGN (Figure \ref{fig:csfandsspgrid}c) overshoots the \citetalias{Kewley01} line due to a harder continuum from younger stars, and much of the grid falls above the real data in the OI plot. Both SFH choices are unrealistic, and ideally we would use a combined continuous + bursty SFH to model a realistic dwarf. However, bracketing reality using the simplified cases of the CSF and the 20 Myr SSP, we see that the OI plot is always better than the NII plot at identifying AGN in metal-poor star-forming dwarfs. \\

Despite the SFH caveat, we note that our models have better overlap with the data than some widely used standard models in literature \citep[e.g.,][]{Groves04, Levesque10}. The classification scheme we will propose in Section \ref{sec:classify}, however, is unaffected by how well our new models match real data because in what follows, we continue to use the \citetalias{Kewley01} maximum theoretical starburst lines for classification. Nonetheless, in the following sections, we will use our new models to demonstrate the need for a more optimized classification scheme using those same lines, especially for dwarf AGN, and to compute gas-phase metallicities for our sample. \\

\subsection{Metallicity Estimation and Fiducial Dwarf Parameters}\label{subsec:metallicity}
To estimate the metallicities and ionization parameters of our galaxies, we run the Bayesian inference code, NebulaBayes \citep{Thomas18}, with models from a slice of our new photoionization grid with pure star formation (0\% AGN contribution) and a CSF SFH. We defer exploration of the effect of varying the AGN fraction and other parameters given in \citet{Richardson22} to a future paper.  Currently,  we  only  consider  the six SELs present in the three diagnostic plots we are using: H$\beta$ 
, [O~III], [O~I], [N~II], H$\alpha$, and [S~II] doublet. For each galaxy, the code compares observed emission lines to predicted emission lines from the photoionization grid, then calculates the posterior probability of each combination of Z and U given flat priors. After evaluating all combinations of Z and U, the code marginalizes over nuisance parameters and obtains the best estimates for Z and U from the 2-D joint probability distribution functions. We use the Bayesian-inferred metallicity estimates to explore the physical properties of the newly identified dwarf AGN in Section \ref{sec:physical}. Based on the Bayesian estimates, we define a fiducial dwarf as one having Z=0.4 \zsun and log(U)=$-$3.25, the median values of Z and U for the combined RESOLVE and ECO SEL sample.  \\

\section{Optimized Emission Line Diagnostic Classification}\label{sec:classify}

\begin{figure*}[]
    \centering
    \includegraphics[width=\linewidth]{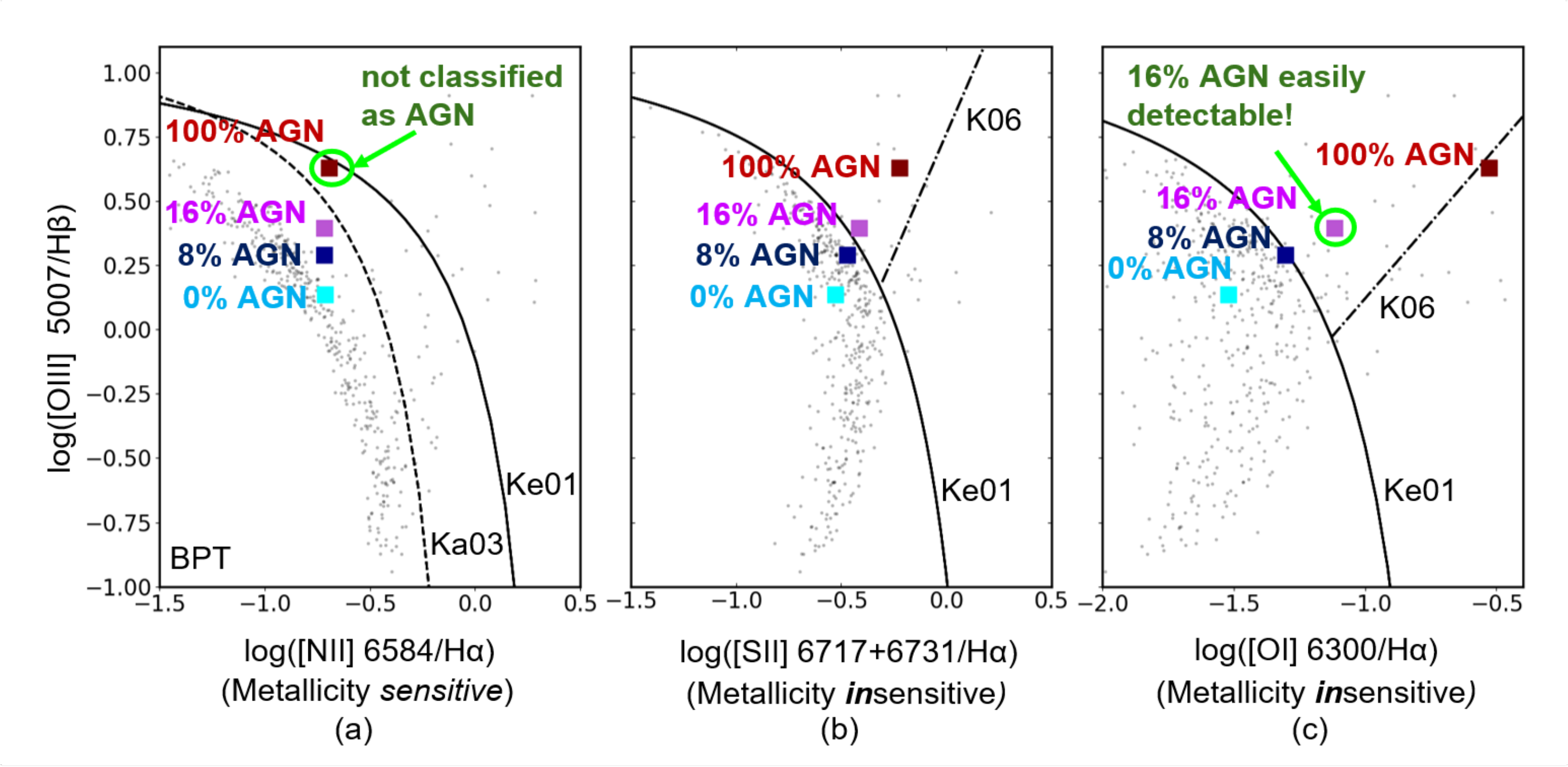}
    \caption{Diagnostic plots showing RESOLVE galaxies (light gray) over-plotted with model data points (cyan, dark blue, purple, brown) for a low-metallicity theoretical dwarf galaxy with a CSF history, a metallicity of Z=0.4Z$_\odot$, and a 0, 8, 16, or 100$\%$ AGN contribution to its spectrum from a BH with \mbh = $10^5$ \msun. The solid line in each plot represents the theoretical maximum starburst line as given by \citetalias{Kewley01}, the dashed line represents the Composite line as given by \citetalias{Kauffmann03}, and the dot-dashed line represents the Seyfert/LINER dividing line as given in \citet[][K06]{Kewley2006}. Galaxies above the \citetalias{Kewley01} lines in all three plots are classified as `Traditional AGN' (Seyferts and LINERs; see Section \ref{sec:classify}), and those between the \citetalias{Kewley01} and \citetalias{Kauffmann03} lines in the \niiplot (BPT; panel (a)) are classified as `Composite.' The \niiplot does not identify even the 100$\%$ AGN model due to its bias against low-metallicity AGN. The \siiandoiplot do much better at identifying dwarf AGN due to their metallicity insensitivity. The \oiplot can identify AGN with spectral contributions almost as low as 8\%, making it the best suited for finding AGN in highly star-forming dwarfs.}
    \label{fig:gridsimple}
\end{figure*}
\begin{figure*}[]
    \centering
    \includegraphics[width=\linewidth]{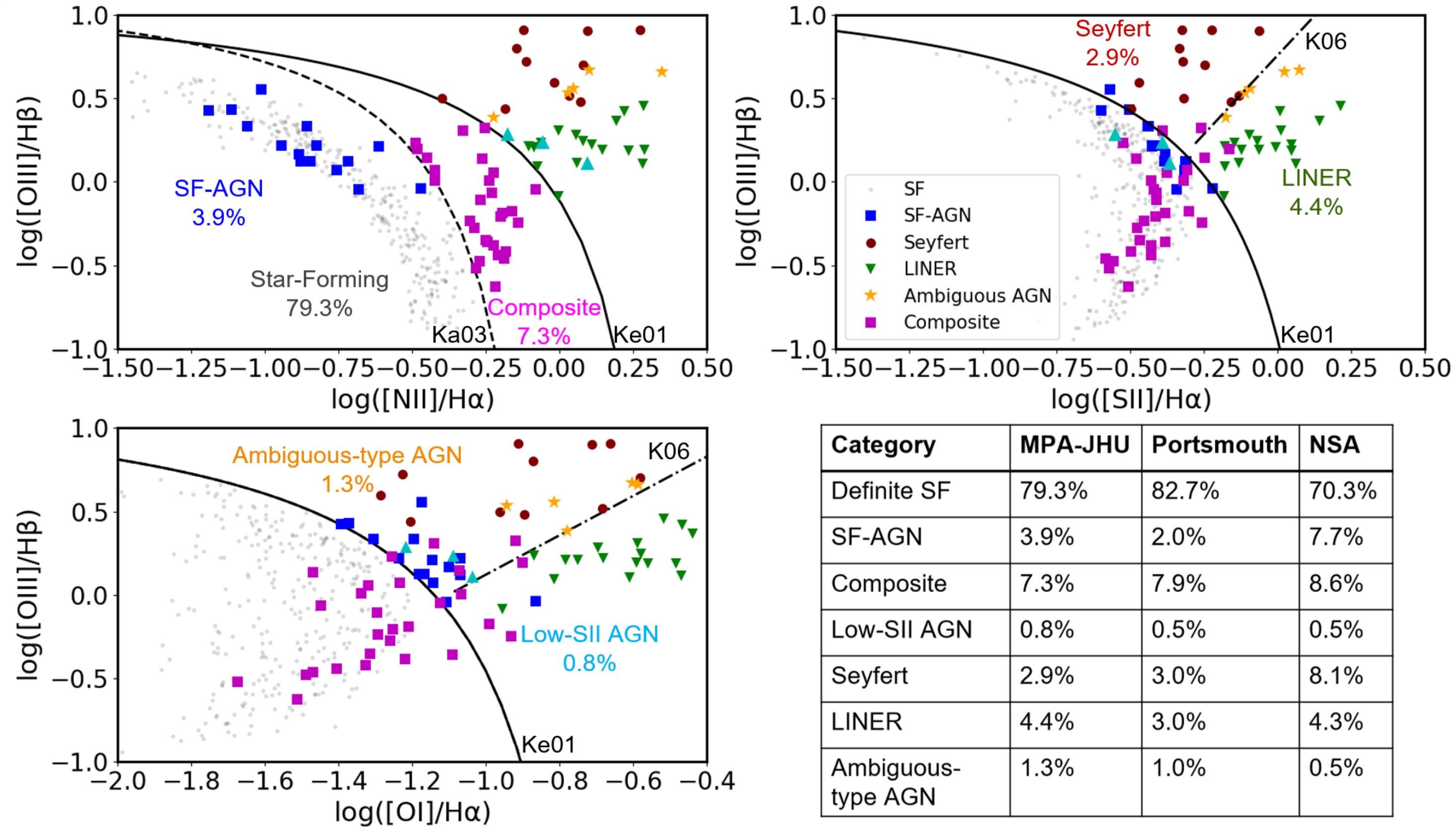}
    \caption{Optimized classification scheme based on NII \citep[BPT;][]{Baldwin81}, SII, and OI \citep{Veilleux87} optical diagnostic plots, shown using fluxes from the SDSS MPA-JHU catalog for galaxies in the RESOLVE survey (see Section \ref{sec:data}). Demarcation lines are the same as in Figure \ref{fig:gridsimple}. To account for differential classifications among the three plots, our optimized scheme introduces two new categories: (i) SF-AGN (blue squares), which are classified as SF by the \niiplot but as AGN (Seyfert or LINER) by the SII and/or OI diagnostic plots, and (ii) Low-SII AGN (cyan triangles) or Low-OI/Low-SII+OI AGN (nonexistent in ECO/RESOLVE), which are classified as AGN (Seyfert or LINER) by the \niiplot but as SF by the SII and/or OI diagnostic plots. Other symbols are described in the legend in the upper right panel. All categories in this optimized scheme are mutually exclusive of each other as described in Section \ref{subsec:newscheme}. The statistics of all three SDSS catalog samples are given in the inset table, and the MPA-JHU statistics are also shown on the plots (see Section \ref{subsec:agnstats} for discussion).}
    \label{fig:classification}
\end{figure*}
We aim to build a classification scheme that can robustly identify AGN in dwarfs and giants alike. In this section, we demonstrate the need for a new classification scheme, and we describe this new optimized scheme.

\subsection{Need for a New Classification Scheme}\label{subsec:need}
The NII plot alone is an inadequate classifier because its abscissa, \NvH, is a crude proxy for metallicity. Typical metal-poor dwarfs have relatively high \OvHb emission and relatively low \NvH \citep{Moustakas06}, usually placing them the left side of the plot below the demarcation line from \citet{Kauffmann03}. Thus, AGN in low-metallicity dwarfs are likely placed in the star forming region of the NII plot \citep{Groves06, Ludwig12}. \\

In Section \ref{subsec:newscheme}, we will present a new classification scheme that optimally uses the NII, SII, and OI plots together to classify every galaxy uniquely. The OI plot is not commonly used since [OI] is a weaker emission line than [NII], but in this section, we show the value of using this less metallicity-sensitive plot as well as the SII plot. In Figure \ref{fig:csfandsspgrid}, our models show that the metallicity-sensitive NII plot does not identify even 100$\%$ AGN contributions to the spectra of metal-poor galaxies, regardless of their SFHs. For a fiducial dwarf representing the RESOLVE and ECO SEL sample with Z = 0.4 \zsun (lime-green star; see Section \ref{subsec:metallicity}), even with a 100\% AGN contribution, it falls below the traditional AGN demarcation line in the NII plot, instead landing in the ``Composite" SF+AGN region (between the dashed and solid curves in Figure \ref{fig:csfandsspgrid}). A study by \citet{Groves06} also finds that the model grids for low metallicity AGN and low metallicity starbursts overlap on the NII plot. In contrast, the OI plot easily identifies a 100\% AGN contribution for all metallicities due to its metallicity insensitivity, regardless of the SFH. \\

Figure \ref{fig:gridsimple} shows that a simulated fiducial \zzero dwarf AGN with low metallicity (Z = 0.4 \zsun) is not classified as an AGN by the NII plot but is easily identifiable as an AGN by the SII and OI plots, which are much less sensitive to metallicity. Figure \ref{fig:gridsimple} also shows that the SII and OI plots outperform the NII plot at identifying AGN with low (8-16\%) spectral contributions in a fiducial metal-poor dwarf. This fiducial dwarf is modelled with a CSF SFH, but the same result stands for a dwarf with an SSP SFH. Dwarfs are expected to have central BH masses ${\sim}10^3{-}10^5$ \msun \citep{Greene20}. At such low masses, even if these BHs were accreting at the Eddington limit, they would be quite faint in optical wavelengths \citep{Greene07}. Even without intense star formation, these faint IMBH signatures may be diluted by the host galaxy \citep{Moran02}, so in a highly star-forming galaxy like a \zzero dwarf, BH signatures are  easily masked by much stronger star formation signatures \citep{Reines13}. Nonetheless, the OI plot can easily identify 16\% AGN spectral contributions. It can even identify AGN contributions down to almost 8\%, making it well suited to finding AGN in highly star-forming galaxies like typical \zzero dwarfs (Figure \ref{fig:gridsimple}). The SII plot is also better than the NII plot at identifying AGN in metal-poor and/or star-forming galaxies, albeit it is more sensitive to star formation dilution than the OI plot. \\

Previous studies have also used the NII, SII, and OI plots together for spectral classification. \citet[][hereafter K06]{Kewley2006} designed a method of using the NII, SII, and OI plots to classify galaxies as Star Forming HII regions (abbreviated as SF), Composite, Seyfert, and LINER galaxies. However, their method was not optimized for dwarfs, and it does not robustly classify all the galaxies in a sample. The \citetalias{Kewley2006} scheme marks galaxies as `Ambiguous' either if they are classified differently as Seyfert vs. LINER by the SII and OI plots, or if they are classified as `Composite' by the NII plot and as Seyfert/LINER in the SII and/or OI plots. It is unclear whether these `Ambiguous' galaxies are always treated as AGN candidates in the rest of the \citetalias{Kewley2006} analysis. Additionally, there is no classification that explicitly includes galaxies classified differently as SF by the NII plot and as AGN by the SII and/or OI plots or vice-versa. These galaxies, as Figure \ref{fig:gridsimple} shows, may include many dwarf AGN. \\

\subsection{New Optimized Scheme}\label{subsec:newscheme}
Our optimized classification scheme assigns a unique category to every galaxy to create a more systematic classification scheme that does not exclude any galaxies, as shown in Figure \ref{fig:classification} using the RESOLVE sample. Following the convention of \citetalias{Kewley2006}, the SII plot uses the sum of the doublet [S~II] $\lambda$6717 + [S~II] $\lambda$6731 (i.e., [S~II] $\lambda$6720), and the NII plot uses only the [N~II] $\lambda$6584 line flux. However, to enhance S/N, we combine the [N~II] $\lambda$6548 and $\lambda$6584 doublet fluxes and then scale them back to get the [N~II] $\lambda$6584 flux given that the ratio of [N~II] $\lambda$6584 to [N~II] $\lambda$6548 is approximately 3:1 \citep{Acker89}. \\

\begin{figure*}[]
    \centering
    \includegraphics[width=\linewidth]{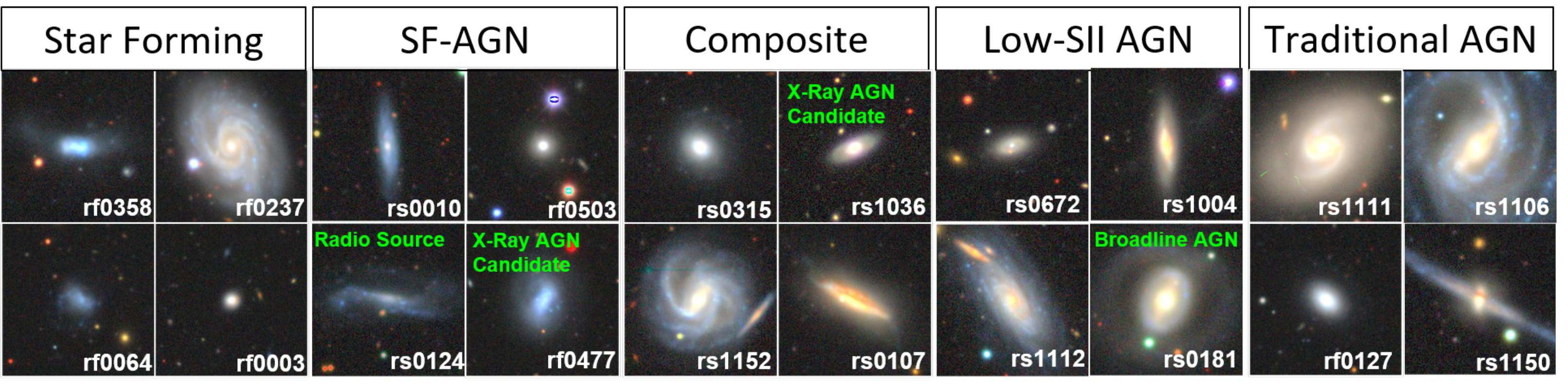}
    \caption{DECaLS DR8 images of representative RESOLVE galaxies in different categories of our optimized emission line classification scheme (Section \ref{subsec:newscheme}). Galaxies with non-optical counterparts or previously catalogued broad-line emission are labelled with green text.}
    \label{fig:class_pics}
\end{figure*}

The theoretical maximum starburst lines for the NII, SII, and OI plots are defined by equations 5, 6, and 7 from \citet{Kewley01} and we refer to them as \citetalias{Kewley01} lines in the respective plots. The Composite line is given by equation 1 in \citet{Kauffmann03}, which we refer to as the \citetalias{Kauffmann03} line hereafter. The dividing lines between Seyfert and LINER galaxies are given by equations 9 and 10 for the SII and OI plots respectively in \citetalias{Kewley2006}, so we hereafter call these the \citetalias{Kewley2006} Seyfert/LINER dividing lines. Below we detail the mutually exclusive categories in our optimized classification scheme. Figure \ref{fig:class_pics} shows representative images of galaxies from the RESOLVE survey in the categories of our optimized classification scheme. We note that we do not explicitly account for the errors on the flux ratios while classifying galaxies in this scheme, similar to general practice in the literature. Our strict S/N cuts minimize error bars, and we have verified that any changes in statistics due to shifting classification of galaxies close to the dividing lines are within the reported errors on the classification percentages.

\begin{enumerate}
    \item \textbf{SF or Definite Star-Forming Galaxies:} These are galaxies that lie below the \citetalias{Kauffmann03} Composite line in the NII plot and to the left of the \citetalias{Kewley01} theoretical maximum starburst line in both the SII and OI plots.
    
    \item \textbf{Composite Galaxies:} These are galaxies that lie between the \citetalias{Kewley01} and the \citetalias{Kauffmann03} lines in the NII plot, regardless of location in the OI or SII plots, unlike \citetalias{Kewley2006}. 

    \item \textbf{Seyfert Galaxies:} These are galaxies that lie above the \citetalias{Kewley01} theoretical maximum starburst lines in the NII, SII, and OI plots, and above the \citetalias{Kewley2006} Seyfert/LINER dividing lines in the SII and OI plots.

    \item \textbf{LINERs:} These are galaxies that lie above the \citetalias{Kewley01} theoretical maximum starburst lines in the NII, SII, and OI plots and below the \citetalias{Kewley2006} Seyfert/LINER lines in the SII and OI plots.
    
    \item \textbf{Ambiguous-type AGN Galaxies:} These are galaxies that lie above the \citetalias{Kewley01} line in the NII plot, but have different AGN types in the SII and OI plots -- Seyfert in (either) one and LINER in the other. Our `Ambiguous-type' galaxies are a subset of the `Ambiguous' galaxies defined by \citetalias{Kewley2006} (see Section \ref{subsec:need} for discussion of the \citetalias{Kewley2006} definition). 
    
    \item \textbf{SF-AGN Galaxies:} These are galaxies that are classified as SF (below \citetalias{Kauffmann03} line) in the NII plot, but as Seyfert or LINER in the SII and/or OI plots (above \citetalias{Kewley01} lines). We will show that these galaxies are mostly metal-poor, gas-rich, and highly star-forming dwarfs that likely host AGN (see Sections \ref{sec:identify} and \ref{sec:physical}).   
    
    \item \textbf{Low-SII, Low-OI, or Low-SII+OI AGN Galaxies:} These are galaxies that are classified as AGN in the NII plot (above \citetalias{Kewley01} line), but as SF in either the SII or OI plots or both (below \citetalias{Kewley01} lines). In our sample, all galaxies in this category are AGN in the NII and OI plots, but SF only in the SII plot, i.e., they are Low-SII AGN. Our sample does not contain any low-OI AGN or Low-SII+OI AGN, although examples could exist in a larger sample especially given observational errors. In the rest of this paper, we will refer to this category as Low-SII AGN. We suspect that the SF classification from the SII plot is spurious and that these galaxies are truly AGN hosts (see below).  
    
\end{enumerate}

The above scheme has introduced two new categories of galaxies -- SF-AGN and Low-SII AGN.  Most of the new AGN candidates we identify are in the SF-AGN category (see inset table of Figure \ref{fig:classification} and Section \ref{subsec:agnstats}). Regardless of the SDSS catalog used, $<$1\% of RESOLVE and ECO galaxies fall in the Low-SII AGN category. As mentioned above, all Low-SII AGN galaxies in our sample have AGN-like (Seyfert or LINER) line ratios in the NII and OI plots, but not in the SII plot. All Low-SII AGN galaxies have high mass and high metallicity, except for one galaxy, \textit{rs0672}. The high metallicity can boost \NvH but not \SvH. We do not have evidence of high SF for the Low-SII AGN hosts, but we also do not have detailed enough information to determine whether SF dilution can cause the differential classifications for these galaxies. Factors like dust and metallicity may drive these differential classifications, but it is beyond the scope of the data we have to disentangle these factors. Images from DECaLS DR8\footnote{www.legacysurvey.org/dr8} (Figure \ref{fig:classification}) also confirm that all of these galaxies have bright nuclei -- \textit{rs0181} even has a broad \ha feature -- suggestive of real AGN activity. Since the OI plot is a more sensitive indicator of AGN than the SII plot, and it corroborates the AGN classification from the NII plot, we choose to consider Low-SII AGN galaxies as AGN candidates. \\

Based on the classifications of RESOLVE and ECO SEL galaxies, we find that the SII plot identifies far fewer dwarf AGN compared to the OI plot. This result is likely because the SII plot cannot identify AGN spectral contributions lower than $\sim$16\%, while  the  OI  plot  identifies contributions down to almost 8\%, as seen in Figure \ref{fig:gridsimple}. \citealt{Richardson22} also shows that \SvH is less sensitive to low AGN contributions than \OvH. However, if the OI plot cannot be used due to practical/observational constraints, the SII plot can still be valuable and recover some AGN in metal-poor and/or star-forming galaxies that would be missed by the NII plot. \\

In the rest of this paper, we will not distinguish between the Seyfert, LINER and Ambiguous-type AGN categories. We consider all of these to be traditionally identified AGN as they lie above the \citetalias{Kewley01} line in all three plots, and we call them ``Traditional AGN'' hosts. Even though there is some contention about the true nature of LINERs, there is still substantial evidence that they are some flavour of AGN \citep{Ho03, Kewley2006, Ho08,  Goulding09}. We count all galaxies in the SF-AGN, Composite, Traditional AGN, and Low-SII AGN categories as AGN candidates. Throughout this paper, we use the terms AGN and AGN candidates interchangeably since no AGN detection method is completely foolproof by itself. \\

\section{Identifying AGN Candidates at \zzero}\label{sec:identify} 
Having used our optimized classification scheme to identify AGN candidates in the RESOLVE and ECO surveys, we will now argue that SF-AGN indeed represent a population of AGN that has not been counted before. We will also analyze the statistics and properties of these AGN and examine how these results depend on differing spectral modelling choices and selection biases between the three SDSS catalogs. \\

\subsection{Are SF-AGN really AGN?} \label{subsec:real}
Several other mechanisms produce AGN-like line ratios in galaxies and could be mistaken for AGN activity. Here we explore such mechanisms and describe why they likely do not explain the behaviour of SF-AGN galaxies. We also present ancillary data to argue that SF-AGN are indeed AGN candidates. 
\begin{enumerate}
    \item \textbf{Could SF-AGN be due to hard radiation from stars?} \\
    Theoretically, extreme starbursts can lead to emission-line signatures similar to those from AGN. In Figure \ref{fig:csfandsspgrid}, our photoionization model with an SSP viewed at 20 Myr shows that AGN-like \OvH ratios can be observed even with 0\% AGN spectral contribution, especially for galaxies with high metallicity and high ionization parameter. The hardening of spectra from young starbursts can potentially be explained by the presence of Wolf-Rayet (W-R) stars within the lower metallicity regime of our galaxies \citep[e.g.,][]{Brinchmann08}. Extremely starbursting galaxies like blue compact dwarfs or blue nuggets have high specific SFRs (sSFRs) generally with log(sSFR) [Gyr$^{-1}$] $> -0.5$ \citep[e.g.,][]{Hopkins02} and high equivalent widths for the \ha line ($>$80$\AA$, \citealt{Lee09}). From Table 2 of \citet{Lee09}, we estimate their median log(sSFR) [Gyr$^{-1}$] to be $\sim$-0.18, meaning their starburst definition (\ha EW$ > $80$\AA$) yields more extreme starbursts than \citet{Hopkins02}. The SF-AGN in our samples have sSFRs (computed from global photometry reflecting 100  Myr timescales; see Sections \ref{subsec:surveydata} and \ref{subsec:sfr} for details on SFRs) comparable to starbursts with a median of log(sSFR) [Gyr$^{-1}$] $\sim -0.27$, but only $\sim$7\% of SF-AGN have EW(\ha) $>$ 80$\AA$ (computed from SDSS fiber spectroscopy).  Given that our global SFRs are on average slightly higher than those from \citet{Salim16} (see Section \ref{subsec:surveydata}), these results do not suggest we are underestimating the role of SF in producing \OvH in SF-AGN. \\
    
    For a more direct comparison on the nuclear scale of the SDSS fibers, we have checked the properties of SF-AGN against the largest local W-R galaxy database compiled by \citet[][henceforth B08]{Brinchmann08}. Figure \ref{fig:oi_oiii} shows the emission line ratio trends of SF-AGN and \citetalias{Brinchmann08} W-R galaxies in a different diagnostic plot whose x-axis is a ratio of neutral-to-doubly-ionized oxygen. \citetalias{Brinchmann08} found that W-R galaxies were offset from the rest of the sample towards the bottom left of the diagnostic plot (see Figure 11 in \citetalias{Brinchmann08}). We see that only a small number of SF-AGN ($<$10\%) with low [O~I]/[O~III] are found in the region populated by W-R galaxies. In fact, in this diagnostic, most SF-AGN have emission line trends that are consistent with Traditional AGN and Composites. Of course, we cannot rule out the coexistence of starbursting nuclear star clusters (NSCs) and AGN, both of which would be much smaller than the SDSS fiber. To assess possible mixtures, we have performed preliminary analysis of AGN spectral contributions using NebulaBayes with our most starbursting photoionization model (20Myr old SSP). We find that $\sim$92\% of SF-AGN require a non-zero AGN contribution to their spectra to explain their observed emission lines. \\
    
    \begin{figure}
    \centering
    \includegraphics[width=\linewidth]{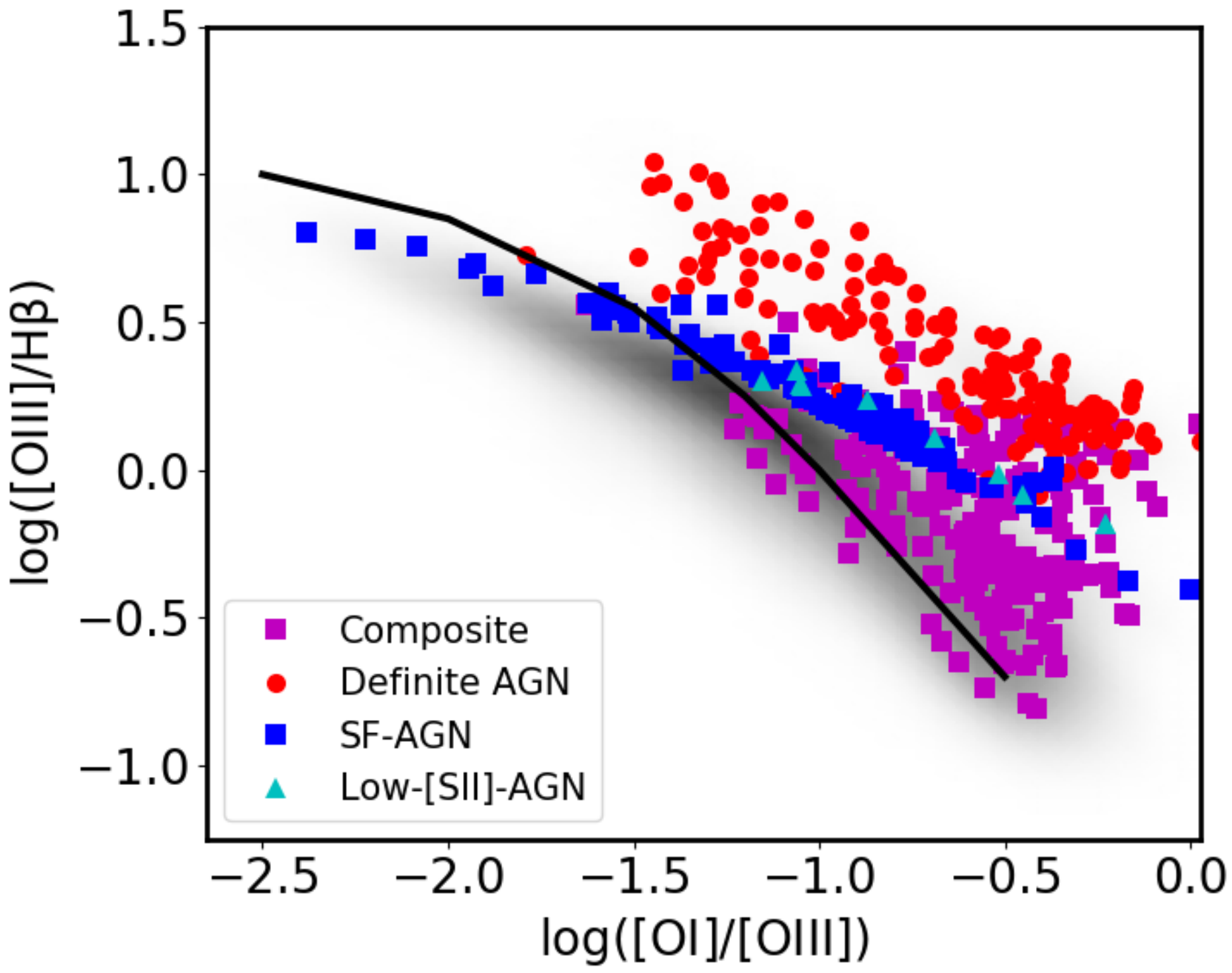}
    \caption{Diagnostic plot of \OvHb vs [O~I]/[O~III] for the combined RESOLVE+ECO SEL sample. Blue squares are SF-AGN, magenta squares are Composites, cyan triangles are Low-SII AGN, and red circles are Traditional AGN (union of Seyfert, LINER and Ambiguous-type AGN categories; see Section \ref{subsec:newscheme}). Black line shows the upper locus of W-R galaxies from \citet{Brinchmann08}. Greater than 90\% of SF-AGN are positioned above the W-R locus in this diagnostic plot, indicating that most SF-AGN cannot be explained by W-R emission alone.}
    \label{fig:oi_oiii}
    \end{figure}

    High Mass XRBs(HMXBs) found in the presence of extreme star-formation can also harden ionizing radiation causing elevated [O~I] fluxes. \citet{Lehmer21} find that HMXB spectral contributions are expected to increase with decreasing metallicity for galaxies with SFRs spanning a wide range from 0.01 to 100 \msun/yr. \citet{Senchyna20} study 11 metal-poor star-forming dwarfs (0.05 - 0.35 \zsun) with nebular He II indicating the presence of high energy photons. They compare the dwarfs with photoionization grids with simple blackbody SEDs to model HMXBs and find little significant contribution of high energy photons from HMXBs. In contrast, \citet{Simmonds21} use photoionization grids with more realistic SEDs and find that HMXBs in low metallicity ($<$0.2 \zsun) dwarfs can potentially elevate \OvH if a specific SED is assumed. The new SF-AGN in this work have metallicities between 0.3 and 0.4 \zsun, higher than those of the \citet{Senchyna20} sample (with a similar range of SFRs) and also higher than the \citet{Simmonds21} models. We have a X-ray luminosity for only one SF-AGN, which has a lower L$\rm_{X-ray}/$SFR ratio than the model analysed in \citet{Simmonds21}; all other SF-AGN are either not targeted or not detected by Chandra or X-ray Multi-Mirror Mission (XMM; see \#3). Based on their metallicity range and lack of exceptionally high X-ray luminosities, we conclude that SF-AGN are unlikely to have elevated [O~I] fluxes solely due to HMXB-hardened spectra. \\

    In summary, the emission line ratios and trends for a majority of SF-AGN are difficult to explain without the presence of AGN activity, even assuming extreme star formation.   
    
    \item \textbf{Do we have evidence of ``nuclear" activity in SF-AGN?}\\ 
\begin{figure}
    \centering
    \includegraphics[width=\linewidth]{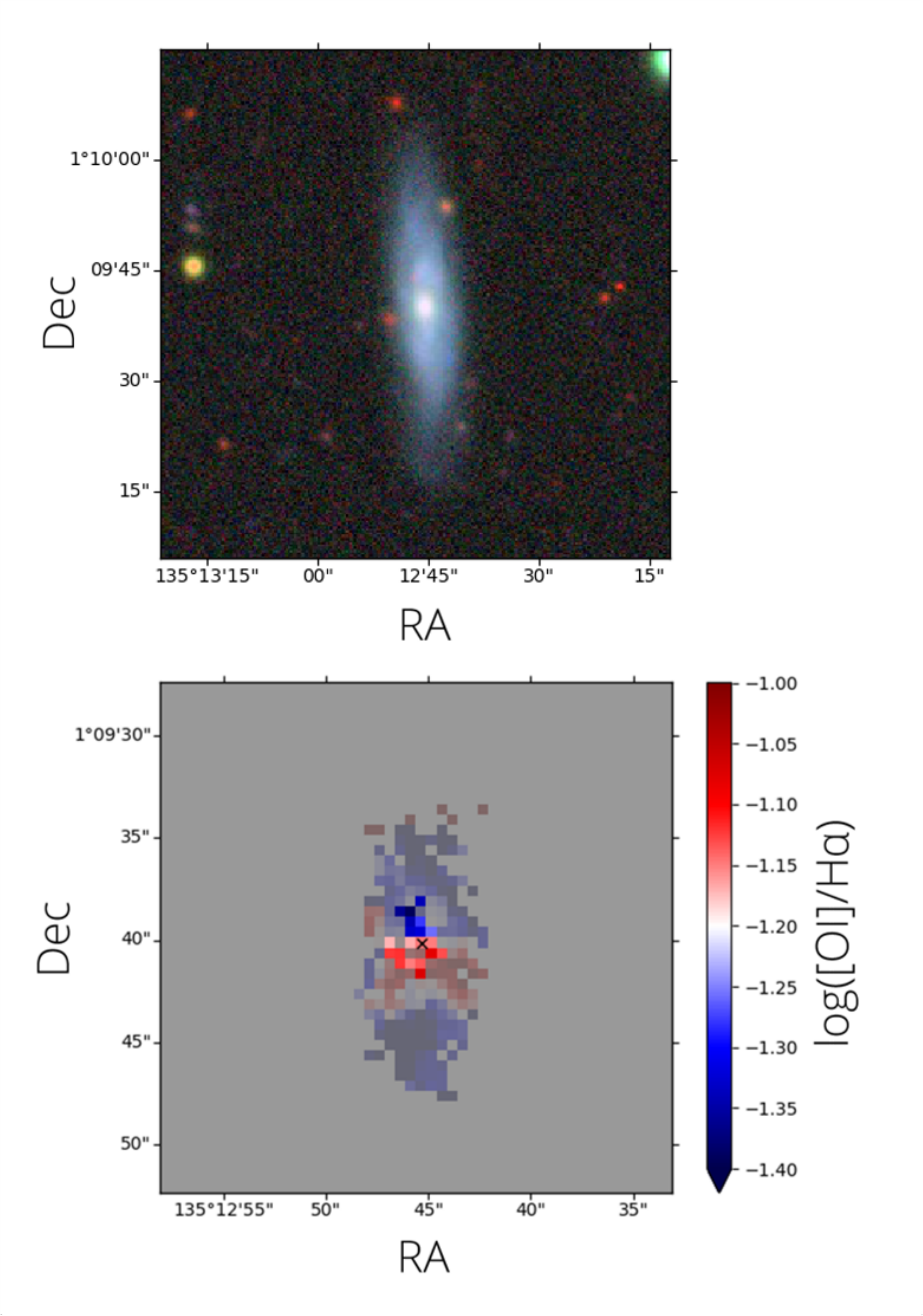}
    \caption{Image and IFU data for a SF-AGN. Upper panel: DECaLS \textit {grz} image of \textit{rs0010}, $\sim$1$'$ across. Lower panel: SAMI map of the central $25''$. Colors represent log(\OvH), which is red above $-1.2$ (AGN-like, see Figure \ref{fig:rs0010bpt}). The brightly colored spaxels in the center have SEL S/N $> 5$ and continuum S/N $ > 10$. The faded and grey spaxels do not meet the emission and/or continuum S/N criteria.}
    \label{fig:rs0010}
\end{figure}
\begin{figure*}
    \centering
    \includegraphics[width=\linewidth]{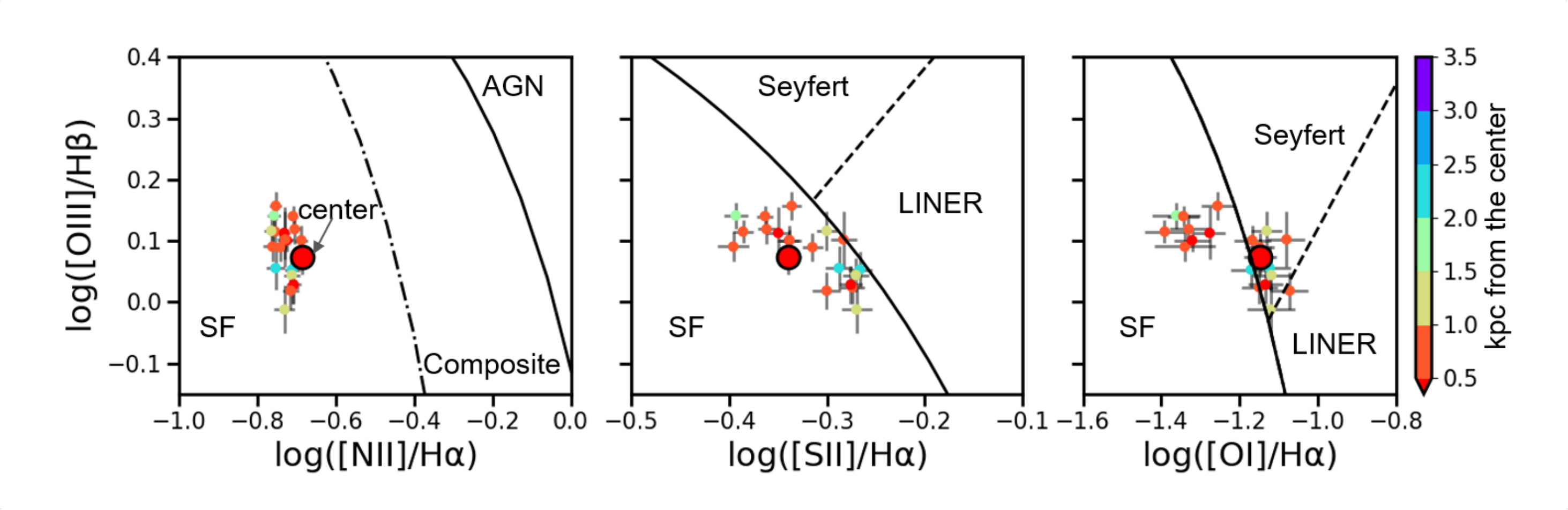}
    \caption{Spatially resolved diagnostic plots using high signal-to-noise SAMI DR2 IFU data for SF-AGN \textit{rs0010}. AGN-like \OvH ratios occur in  several spaxels within 2 kpc of the center.}
    \label{fig:rs0010bpt}
\end{figure*}    
    A crossmatch of all unique RESOLVE and ECO SEL galaxies with the SAMI DR2 catalog \citep{Bryant15} yielded integral field unit (IFU) data for two SF-AGN. For each IFU data cube, we filter spaxels to have both continuum S/N $> 5$ and emission line S/N $ > 5$ for all SELs. This filtering is necessary because a high continuum S/N ensures reliable continuum fitting and signal decomposition, and a high emission line S/N ensures reliable use of the three diagnostic plots. Figure \ref{fig:rs0010} shows spatially resolved high S/N data for \textit{rs0010}, revealing that high AGN-like \OvH ratios (red spaxels) are centrally located and clearly separated from SF-like \OvH ratios (blue spaxels).
    
    Figure \ref{fig:rs0010bpt} shows spatially resolved diagnostic plots for \textit{rs0010}, confirming AGN-like line ratios in the OI plot in several spaxels within the central 2 kpc. Only the OI plot is able to identify the AGN-like line ratios; the NII and SII plots fail, likely due to bias against low metallicity and SF dilution. Another SF-AGN, \textit{rs0775}, shows similar line ratio behaviour, but with worse S/N for all lines, and the S/N of the continuum is much too weak to form any reliable conclusions. In summary, reliable IFU data for the SF-AGN \textit{rs0010} confirm that AGN-like line ratios are centrally located.

    \item \textbf{Do SF-AGN have non-optical or other known counterparts?} \\
        \begin{figure}
        \centering
        \includegraphics[width=\linewidth]{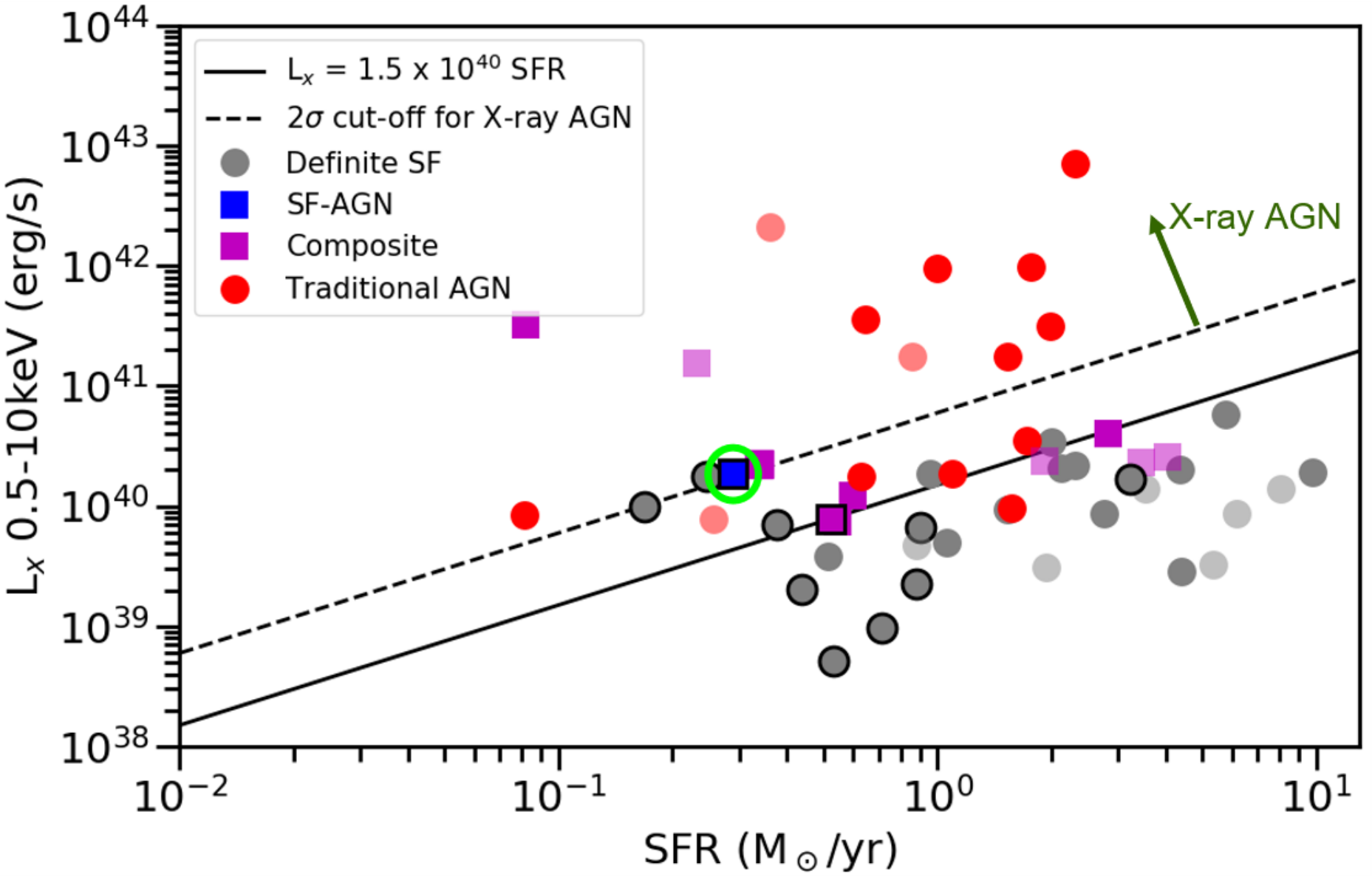}
        \caption{X-Ray luminosity vs. SFR relationship for RESOLVE and ECO SEL galaxies. The darker points represent galaxies with 3XMM fluxes and the lighter points represent galaxies with Chandra fluxes. Points with black outlines are dwarfs. The solid line shows the empirical relationship between the 0.5-10keV X-Ray luminosity and star formation rate given by \citet{Ranalli03}. Galaxies that fall beyond 2$\sigma$ (0.2 dex; dashed line) above the relation likely host AGN. Only one SF-AGN, \textit{rf0477}, borderline qualifies as an X-ray AGN and is highlighted by a green circle. Note: two data points are slightly offset to show the SF-AGN clearly. }
        \label{fig:xray}
    \end{figure}
    Table \ref{table:3} reports the statistics of RESOLVE and ECO SF-AGN dwarfs with non-optical or other known counterparts. We do not find any crossmatches (within a $5''$ radius) of RESOLVE and ECO SF-AGN in two optical AGN catalogs \citep{Veron06, Flesch15} and a comprehensive broadline AGN catalog \citep{Liu19}. \\ 
    
    We have also crossmatched the RESOLVE and ECO SEL samples with the 3XMM-DR8 catalogue \citep{Rosen16} and the Chandra Source Catalog (CSC) Release 2.0 \citep{Evans20}. We find an X-ray match from 3XMM for only one RESOLVE SF-AGN, \textit{rf0477}, and none from the CSC. To assess whether this detection represents an X-ray AGN or simply X-ray binaries, we put the catalog X-ray fluxes on a common basis with those used in the $\rm L_{X-ray} - SFR$ relation of \citet{Ranalli03} as shown in Figure \ref{fig:xray}. We multiply the 0.5-12keV 3XMM flux by a factor of 0.9 (based on a photon index, $\Gamma = $ 1.7) to obtain 0.5-10keV flux \citep{Agostino18}. In Figure \ref{fig:xray}, we also show X-ray crossmatches for RESOLVE and ECO SEL galaxies that are not SF-AGN for reference, including the Composite that is an X-ray candidate shown in Figure \ref{fig:class_pics}. For galaxies with CSC matches, we multiply the 0.5-7.0keV CSC flux by a factor of 1.21 to obtain 0.5-10keV fluxes \citep{LaMassa13}. The SF-AGN \textit{rf0477} borderline qualifies as an X-ray AGN candidate. \\ 
    
    We have also looked for radio counterparts of RESOLVE and ECO SEL galaxies in the HEASARC Master Radio Catalog\footnote{https://heasarc.gsfc.nasa.gov/w3browse/master-catalog/radio.html}, finding crossmatches for 1 ECO SF-AGN and 2 RESOLVE SF-AGN. Based on visual examination of radio continuum cutouts of these three SF-AGN, one galaxy, \textit{rs1038}, shows signs of extended emission, and two galaxies, \textit{rs0124} and \textit{ECO05128}, have unresolved emission. With the available data, we cannot test whether the emission is due to AGN or SF. \\
    
    Finally, we use our own recomputed WISE photometry (Paper II) available for 1324 of the 2605 RESOLVE and ECO SEL galaxies from the MPA-JHU catalog to assess mid-IR color selection criteria for AGN. We consider galaxies as mid-IR AGN candidates if they cross the AGN color threshold using any one of the three widely-used criteria from  \citet{Jarrett11}, \citet{Stern12}, and \citet{Satyapal18}. In practice, the \citet{Jarrett11} criterion is the most restrictive while the \citet{Satyapal18} criterion is the least restrictive. None of the SF-AGN is classified as a mid-IR AGN by any of the criteria.
    
    In summary, among all SF-AGN in the combined RESOLVE and ECO samples, one galaxy has an AGN counterpart from X-rays. There are also 3 SF-AGN with radio crossmatches, albeit the radio emission cannot be classified as having SF or AGN origin with available data. Generally, we conclude that most SF-AGN do not have  counterparts (see Table \ref{table:3}), but this result is not unexpected as most AGN identification techniques are sensitive to finding AGN representing higher metallicity hosts, more massive BHs, and/or AGN with high spectral contributions. Also, the reliability of mid-IR selection for dwarf AGN detection has been debated (see Section \ref{sec:intro}); we investigate this issue further in Paper II.

\begin{table*}[]
\caption{Multi-wavelength cross-match of RESOLVE and ECO SF-AGN dwarfs}
\begin{center}
    \begin{tabular}{C{3cm}|C{3cm}C{3cm}|C{3cm}C{3cm}}
    \toprule
    {} & \multicolumn{2}{c|}{RESOLVE} & \multicolumn{2}{c}{ECO excluding RESOLVE overlap} \\ \hline
    Type & Crossmatches for SF-AGN &  Crossmatched SF-AGN classified as AGN by other methods  &  Crossmatches for SF-AGN &  Crossmatched SF-AGN classified as AGN by other methods   \\
    \hline
    AGN Catalogs\tablenotemark{a}        &   0  &   0    &   0 &  0 \\
    Mid-IR AGN\tablenotemark{b}       &   7 &  0  &    33 &  0 \\
    X-ray AGN\tablenotemark{c} &   1   &  1  &   0 &   0 \\
    Radio Sources\tablenotemark{d} &   2   &   N/A   &   1  &  N/A \\
    \hline
    \end{tabular}%
\end{center}
\tablenotetext{a}{AGN candidates based on $5''$ crossmatch with \citet{Veron06}, \citet{Flesch15}, and \citet{Liu19} catalogs}
\tablenotetext{b}{Union of AGN candidates identified by mid-IR color criteria from \citet{Jarrett11}, \citet{Stern12}, and \citet{Satyapal18}}
\tablenotetext{c}{AGN candidates based on $L_{X-ray}$-SFR relationship \citep{Ranalli03}}
\tablenotetext{d}{Radio sources with unknown AGN status based on $5''$ crossmatch with HEASARC Radio Master catalog}
\label{table:3}
\end{table*} 
    \item \textbf{Could SF-AGN be galaxies with shocks?}\\
    Shocks can cause enhanced optical emission line ratios and are found in galaxies with high star formation \citep{Heckman87, Rupke05}, AGN \citep{Cecil02, Rupke11, DAgostino191}, or galaxy mergers \citep{Rich11, Rupke13}. Shocks and AGN line ratios are expected to behave similarly in the NII plot, but shock emission is localized in the LINER region in the SII and OI plots \citep{Allen08, Davies17}. However, SF-AGN mostly avoid the LINER regions of these plots (Figure \ref{fig:identify}). Additionally, as noted under \#3, broad \ha features have not been catalogued for any of our SF-AGN, although this result does not rule out low velocity shocks with  $v < 500 km s^{-1}$ \citep{Reines13}. Recent work \citep[e.g.,][]{DAgostino191,Molina21} has shown that in galaxies with AGN-like narrow emission line ratios, the AGN can be the origin of observed shocks. \\
    
    In summary, if SF-AGN hosts do have low-velocity shocks, they could potentially originate from the AGN, but such shocks cannot easily explain the Seyfert-like line ratios in the SII and OI plots. 

    \item \textbf{Could SF-AGN be DIG galaxies?} \\
    Diffused ionized gas (DIG) is low surface density \ha gas, typically found in the outskirts of face-on galaxies or in the extraplanar regions of disk galaxies. DIG can  comprise up to $\sim60\%$ of the total gas mass \citep{Zhang17, Vogt17, Lacerda18}. DIG can cause elevated ratios of [NII], [SII], and [OI] with respect to \ha \citep{Kaplan16} and can push galaxies to the Composite/AGN side of the NII plot \citep{Zhang17}. The SF-AGN in our sample, by definition, do not have elevated \NvH, but their low metallicity nature could mask any potential \NvH enhancement from DIG. Regardless, the SDSS spectra we use sample only the central $2''$ or $3''$ where the galaxy light should not be DIG dominated. Additionally, DIG does not easily explain the spatial trends observed in Figure \ref{fig:rs0010bpt} where we see AGN-like \OvH line ratios in only the central 2 kpc of a SF-AGN galaxy.     \\
\end{enumerate}

\subsection{AGN statistics in RESOLVE and ECO}\label{subsec:agnstats}
\begin{table*}

\begin{center}
\caption{Statistics of SEL galaxy categories in our optimized emission-line classification scheme}
    \begin{tabular}{l|ccc|ccc|ccc}
    \toprule
    Category & \multicolumn{9}{c}{Classification of All SEL Galaxies (dwarfs + giants)}           \\
    \hline
    & \multicolumn{3}{c|}{MPA-JHU} & \multicolumn{3}{c|}{Portsmouth} & \multicolumn{3}{c}{NSA}           \\
    \hline
    {} & RESOLVE &  ECO &  Overall\tablenotemark{a} & RESOLVE &  ECO &  Overall & RESOLVE &  ECO &  Overall  \\
    
    ($\#$ of galaxies) & (382) & (2507) & (2605) & (202) & (1161) & (1207)& (209) & (1363) & (1411)         \\
    \hline
Definite SF   &        79.3\% &    81.8\% &        81.6\% &         82.7\% &     83.8\% &        83.5\% &        70.3\% &    70.4\% &        70.2\% \\
SF-AGN     &         3.9\% &     4.0\% &         4.0\% &          2.0\% &      3.4\% &          3.4\% &         7.7\% &     8.6\% &         8.5\% \\
Composite     &         7.3\% &     8.0\% &         7.9\% &          7.9\% &      6.8\% &          6.7\% &         8.6\% &    11.4\% &        11.2\% \\
Low-SII AGN     &         0.8\% &     0.4\% &         0.4\% &          0.5\% &      0.8\% &          0.7\% &         0.5\% &     0.4\% &         0.4\% \\
Seyfert       &         2.9\% &     1.8\% &         1.9\% &          3.0\% &      2.4\% &          2.7\% &         8.1\% &     4.3\% &         4.5\%   \\
LINER         &         4.4\% &     3.5\% &         3.7\% &          3.0\% &      2.4\% &          2.6\% &         4.3\% &     4.3\% &         4.4\% \\
Ambiguous-type AGN &         1.3\% &     0.5\% &         0.5\% &          1.0\% &      0.2\% &          0.2\% &         0.5\% &     0.7\% &         0.8\% \\
    \hline
Traditional AGN\tablenotemark{b} &         8.6\% &     5.8\% &         6.1\% &          7.0\% &      5.0\% &          4.7\% &         12.9\% &     9.3\% &         9.7\% \\
All AGN\tablenotemark{c} &         $20.6^{+2.1}_{-2.0}\%$ &     $18.2^{+2.8}_{-2.5}\%$ &         $18.4^{+1.1}_{-1.0}\%$ &          $17.4^{+2.8}_{-2.5}\%$ &      $16.0^{+1.1}_{-1.0}\%$ &          $16.3^{+1.1}_{-1.0}\%$ &         $29.7^{+3.2}_{-3.0}\%$ &     $29.7^{+1.2}_{-1.2}\%$ &         $29.8^{+1.2}_{-1.2}\%$ \\
\toprule
     & \multicolumn{9}{c}{Classification of SEL Dwarf Galaxies (\mstar $< 10^{9.5}$ \msun)}           \\
\hline
    ($\#$ of galaxies) & (226) & (1525) & (1577) & (129) & (749) & (776)& (114) & (738) & (761)         \\
    \hline
Definite SF   &         93.4\% &     93.1\% &         93.2\% &          96.9\% &      94.0\% &          94.1\% &         85.1\% &     84.3\% &         84.4\% \\
SF-AGN     &          5.8\% &      6.2\% &          6.1\% &           2.3\% &       5.1\% &           4.9\% &         13.2\% &     14.6\% &         14.5\% \\
Composite     &          0.4\% &      0.4\% &          0.4\% &           0.8\% &       0.4\% &           0.5\% &          0.0\% &      0.5\% &          0.5\% \\
Low-SII AGN     &          0.4\% &      0.1\% &          0.1\% &           0.0\% &       0.0\% &           0.0\% &          0.0\% &      0.0\% &          0.0\% \\
Seyfert       &          0.0\% &      0.3\% &          0.3\% &           0.0\% &       0.5\% &           0.5\% &          1.8\% &      0.5\% &          0.7\%  \\
LINER         &          0.0\% &      0.0\% &          0.0\% &           0.0\% &       0.0\% &           0.0\% &          0.0\% &      0.0\% &          0.0\% \\
Ambiguous-type AGN &          0.0\% &      0.0\% &          0.0\% &           0.0\% &       0.0\% &           0.0\% &          0.0\% &      0.0\% &          0.0\% \\
\hline
Traditional AGN &         0.0\% &     0.3\% &         0.3\% &          0.0\% &      0.5\% &          0.5\% &         1.8\% &     0.5\% &         0.7\% \\
All AGN &         $6.6^{+1.8}_{-1.5}\%$ &     $6.9^{+0.7}_{-0.6}\%$ &         $6.9^{+0.4}_{-0.4}\%$ &          $3.1^{+1.9}_{-1.2}\%$ &      $6.0^{+0.9}_{-0.8}\%$ &          $5.9^{+0.9}_{-0.8}\%$ &         $15.0^{+3.6}_{-3.0}\%$ &     $15.6^{+1.4}_{-1.3}\%$ &         $15.6^{+1.3}_{-1.3}\%$ \\
\toprule
     & \multicolumn{9}{c}{Full-Population Dwarf SEL AGN Statistics}           \\
\hline
(\# of dwarfs) &   (648) &     (3931) &    (4161)     &          (648) &     (3931) &    (4161)    & (648) &     (3931) &    (4161)    \\
Dwarf AGN &         $2.3^{+0.7}_{-0.5}\%$ &     $2.7^{+0.3}_{-0.2}\%$ &         $2.6^{+0.2}_{-0.2}\%$ &          $0.6^{+0.4}_{-0.2}\%$ &      $1.1^{+0.2}_{-0.2}\%$ &          $1.1^{+0.2}_{-0.2}\%$ &         $2.6^{+0.7}_{-0.6}\%$ &     $3.0^{+0.3}_{-0.2}\%$ &         $2.9^{+0.3}_{-0.2}\%$ \\
\toprule
\end{tabular}
\label{table:fullstats}
\end{center}
\tablenotetext{a}{Overall RESOLVE+ECO sample does not double-count the overlap between the two surveys}
\tablenotetext{b}{Sum total of Seyfert, LINER, and Ambiguous-type AGN categories}
\tablenotetext{c}{Sum total of SF-AGN, Composite, Low-SII AGN, and Traditional AGN categories}
\tablenotetext{}{Note: All error bars are computed using binomial confidence intervals}

\end{table*}

Table \ref{table:fullstats} gives the statistics of RESOLVE and ECO galaxies in our new mutually exclusive categories using fluxes from the three SDSS catalogs. Including the new categories (SF-AGN and Low-SII AGN), the AGN percentage in all \zzero SEL galaxies is $\sim$16-30\% depending on the SDSS catalog used. We note that it is standard in most AGN studies to express AGN percentages relative to the search sample (i.e., not including galaxies excluded during sample selection). We will usually follow that practice and state statistics as a fraction of the number of galaxies in the search sample, i.e., SEL galaxies. We also examine the AGN percentage normalized to the full galaxy population, and in both cases we explicitly specify the population under consideration to avoid confusion. \\

\begin{figure*}[t!]
    \centering
    \includegraphics[width=\linewidth]{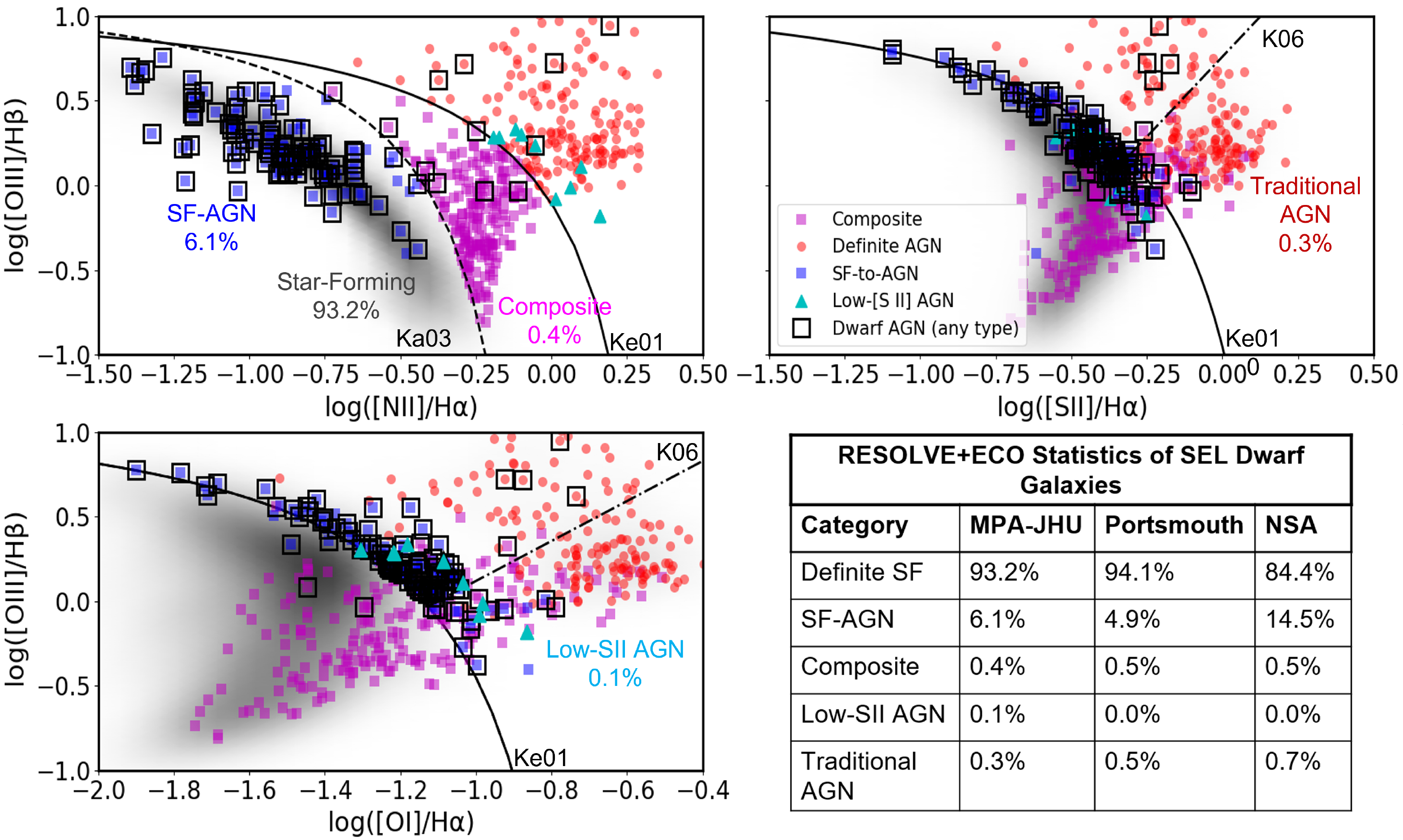}
    \caption{Location of dwarf AGN (black open squares) shown in diagnostic plots for RESOLVE and ECO combined. Demarcation lines are as in Figure \ref{fig:gridsimple}. Grey shading shows the number density of all Star Forming SEL galaxies in the combined RESOLVE and ECO catalogs. Blue squares are SF-AGN: these make up $\sim$75-95\% of dwarf AGN depending on the catalog used. Magenta squares are Composites, cyan triangles are Low-SII AGN, and red circles are Traditional AGN (union of Seyfert, LINER and Ambiguous-type AGN categories). Points without the black squares are not dwarf galaxies. The statistics of all three SDSS catalog samples are given in the inset table, and the MPA-JHU statistics are also shown on the plots (see Section \ref{subsec:agnstats} for discussion).}
    \label{fig:identify}
\end{figure*}

The new SF-AGN category makes up $\sim$3-9$\%$ of the full RESOLVE and ECO SEL samples (i.e., including dwarfs and giants). Most SF-AGN hosts (75\% to 95\% depending on the catalog) are dwarfs. Figure \ref{fig:identify} shows the dwarf AGN candidates in the overall RESOLVE and ECO SEL sample. Most dwarf AGN are in the new SF-AGN category (blue squares). SF-AGN are mainly identified by the OI plot, which is relatively insensitive to both metallicity and SF dilution. On comparison with Figure \ref{fig:gridsimple}, most SF-AGN seem to have AGN spectral contributions in the 8-16\% range. Similar dwarf SF-AGN would have been missed in past studies that use only the NII plot due to their low metallicity as well as their high star formation rates, typical for \zzero dwarfs (see Section \ref{sec:physical}). The previous work of \citetalias{Kewley2006} using all three of the NII, SII, and OI plots likely missed these AGN because none of their categories explicitly includes them. \\

Figure \ref{fig:dwarfpercent} shows that depending on the SDSS catalog used, the new overall dwarf AGN percentage in SEL galaxies is now $\sim$3-16\%, much higher that the $<$1\% in previous studies \citep{Reines13, Sartori15, Reines15}. AGN in dwarf SEL hosts make up $\sim$0.6-3.0\% of the full baryonic mass-limited RESOLVE and ECO samples. The percentage of AGN in SEL giants varies between 36\% to 47\% depending on the sample and catalog. AGN in giant SEL hosts make up around 3-4\% of the full baryonic mass-limited sample. \\

\subsection{Differences in AGN statistics depending on spectral modelling and selection biases}\label{subsec:catalogdiff}
\begin{figure*}[hbt!]
    \centering
    \includegraphics[width=\linewidth]{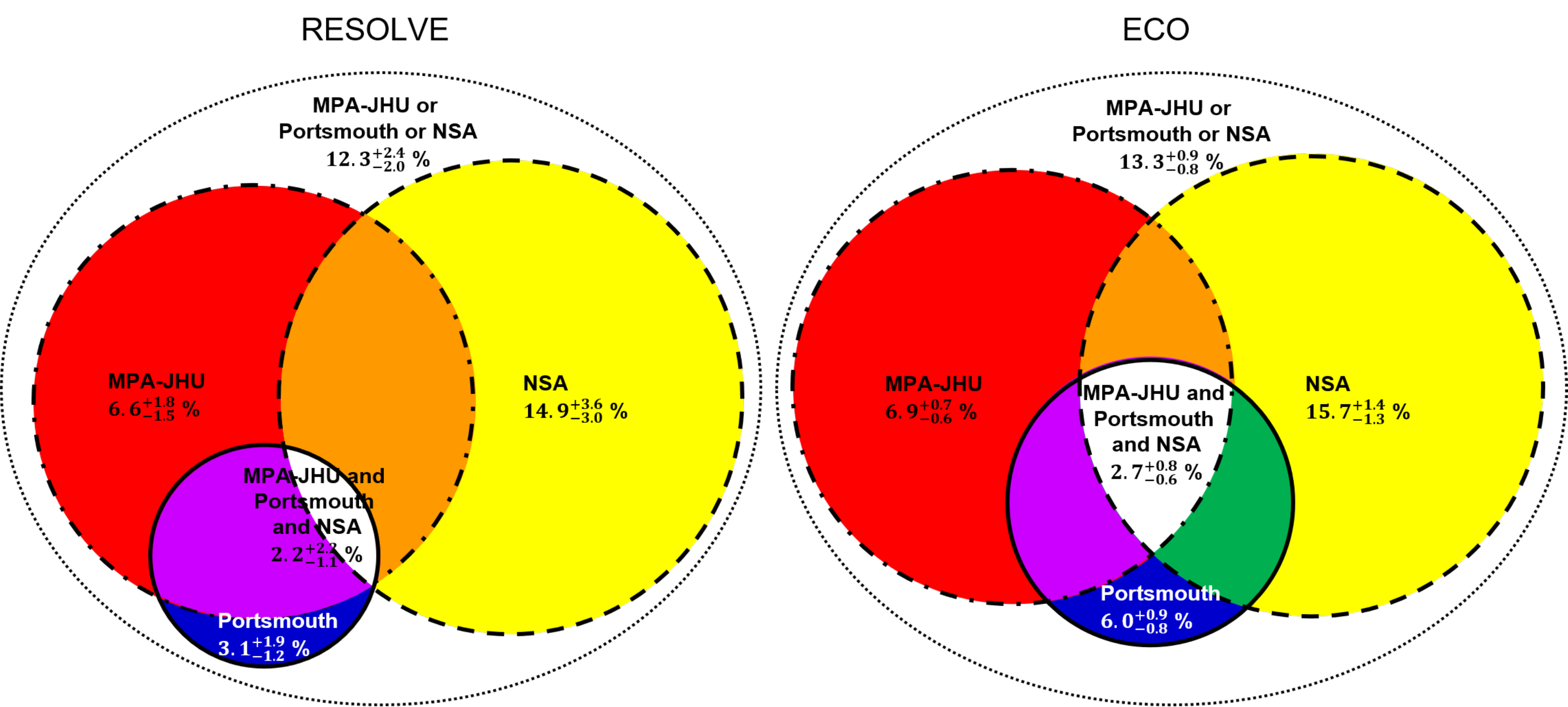}
    \caption{Percentages of dwarf AGN in RESOLVE and ECO SEL galaxies as determined with different catalogs or combinations thereof. Circle areas are proportional to the numbers of dwarf AGN in each sample. Dot-dashed, dashed, and solid circles represent dwarf AGN in the MPA-JHU, NSA, and Portsmouth samples, respectively, while red, yellow, and blue sectors represent dwarf AGN uniquely detected in each of these samples. The white sector represents the intersection of dwarf AGN in all three samples. Big dotted ovals represent the union of all three samples. We consider galaxies in the SF-AGN, Composite, Low-SII AGN, and Traditional AGN categories as candidate dwarf AGN hosts. By including the new categories, the dwarf AGN percentage in RESOLVE and ECO SEL galaxies is now $\sim$3-16\% depending on the catalog used.}
    \label{fig:dwarfpercent}
\end{figure*}
Table~\ref{table:fullstats} shows noticeable variation in the AGN percentages derived from the three different SDSS catalogs. The catalogs have different data sources --- the MPA-JHU and NSA catalogs are based on SDSS DR8, while the Portsmouth catalog is based on SDSS DR12 --- but we do not find evidence that the data source affects the sample selection or the AGN statistics. Cross-matching all RESOLVE and ECO galaxies (including non-EL galaxies), the Portsmouth and MPA-JHU catalogs include exactly the same 7557 galaxies, while the slightly larger NSA catalog has 95\% overlap in galaxies. As shown visually in Figure \ref{fig:dwarfpercent}, less than half as many Portsmouth measurements pass our SEL S/N cuts as do MPA-JHU measurements (16.0\% vs.\ 34.5\%, see Section~\ref{subsec:fluxes}), apparently due to higher error estimates (for our six emission lines of interest, the median Portsmouth errors are 2.5--3.5$\times$ higher than the median MPA-JHU errors). However, the Portsmouth-identified dwarf AGN have substantial overlap with the MPA-JHU-identified dwarf AGN, and the two catalogs yield consistent dwarf AGN percentages within their uncertainties, despite being based on different SDSS Data Releases. In contrast, Figure~\ref{fig:dwarfpercent} shows that the NSA and MPA-JHU catalogs have much lower dwarf AGN overlap and yield discrepant dwarf AGN frequencies, despite being based on the same SDSS Data Release. \\

The choice of stellar population models for spectral decomposition can certainly affect emission line ratios and consequently the sample selection and AGN statistics. The MPA-JHU catalog uses \citet[][hereafter BC03]{Bruzual03} models with varying metallicities to fit the stellar continuum, including a low metallicity model. Both the Portsmouth and NSA catalogs use only solar metallicity models from \citet[][hereafter MS11\textsubscript{solar}]{Maraston11} and try to exploit the age-metallicity degeneracy to model metallicity dependence by using different ages. The authors claim that this method should not greatly affect flux estimates in galaxies with very strong emission lines, like the sample in our study. However, \citet{Reichardt01} find that the age-metallicity degeneracy disappears while simultaneously fitting the absorption lines and the continuum. This may lead to fitting absorption features that are too shallow or too deep, yielding emission fluxes that are too low or too high, as seen in \citet{Chen18}. \\

Apart from the continuum modelling, the MPA-JHU and Portsmouth catalogs rely on similar procedures for extracting fluxes and applying corrections (given our homogenization of extinction corrections, see Section \ref{subsec:fluxes}). For RESOLVE and ECO SEL galaxies that are common to both catalogs, we find a tight correlation between the SEL fluxes with a low spread in values. This trend has also been observed by previous studies that have compared the MPA-JHU and Portsmouth fluxes \citep[e.g.,][]{Thomas13, Chen18, Zaw19}. However, despite the tight correlation between the fluxes, small variations in the flux ratios are enough to change the AGN vs. SF classification of galaxies close to the demarcation lines. Such small variations, whether due to modeling differences or different SDSS data releases, may explain the imperfect overlap between dwarf SEL AGN in the Portsmouth and MPA-JHU catalogs (Figure \ref{fig:dwarfpercent}). \\

AGN statistics are also dependent on the calibrations applied to the datasets. The NSA catalog has an additional flux calibration to fix small-scale calibration residuals that arise from the standard SDSS pipeline \citep{Yan11}. With this reprocessed spectroscopy, the NSA fluxes for RESOLVE and ECO SEL galaxies are on average $\sim$30\% higher than corresponding MPA-JHU or Portsmouth fluxes, and the NSA flux ratios are $\sim$10\% higher. With NSA fluxes, the dwarf AGN percentage among SEL galaxies comes out $\sim$3$\times$ higher than with MPA-JHU or Portsmouth fluxes (see Figure \ref{fig:dwarfpercent}). This difference is somewhat expected considering that \citet{Yan11} apply the NSA flux calibrations to the MPA-JHU catalog, and they find that $\sim$7\% of LINERs have different classifications in the \citetalias{Kewley2006} system using MPA-JHU fluxes with the additional flux calibrations versus without them, indicating that the flux differences are non-negligible. \\

Another likely contributor to the higher dwarf AGN frequencies derived from the NSA vs.\ MPA-JHU or Portsmouth catalogs may be selection effects. The emission line measurements from the three SDSS catalogs are affected by differences in continuum fitting methodology, and our SEL samples are selected based on S/N cuts on these measurements, consequently inheriting selection effects related to the differences in fitting methodology. Comparing the properties of SEL galaxies in the three catalogs (baryonic mass, halo mass and (u-r) color), we see a consistent trend where the NSA SEL catalog has relatively fewer blue-sequence dwarfs in lower-mass halos and more red-sequence giants in higher-mass halos compared to the other two catalogs (Figure \ref{fig:resolvesample_hist} left panel). We suspect that this trend may be because the ratio of NSA catalog errors to MPA-JHU catalog errors is negatively correlated with mass for [N~II] (The NSA errors are also, on average, $\sim$2-3$\times$ larger than the MPA-JHU errors for all emission lines, but this fact is also true for Portsmouth errors, which shows no mass-dependent trends relative to MPA-JHU errors). The NSA errors for [N~II] are $\sim$3$\times$ the MPA-JHU errors on the low mass end but $\sim$0.5$\times$ the MPA-JHU errors on the high mass end. This mass-dependency of errors and therefore S/N likely results in fewer dwarfs passing our SEL S/N cuts. We cannot speculate as to why this trend in errors arises between the catalogs, but it may contribute to the higher dwarf SEL AGN percentage estimated from the NSA catalog -- 16\% compared to 3-6\% from the JHU and Portsmouth catalogs -- if  it reflects modeling differences that preferentially lead to rejecting non-AGN dwarfs from the NSA SEL sample. \\

In summary, the differences we see in AGN percentages for SEL dwarfs seem to be primarily due to irreducible (as per our current knowledge) systematics due to different choices in spectral processing methodologies. \citet{Chen18} and \citet{Zaw19} explore these differences and find that modelling choices dramatically affect AGN categorization. Using only the NII plot, they find that BC03-based fluxes from the MPA-JHU catalog identify more AGN than MS11$\rm_{solar}$-based fluxes from the Portsmouth catalog. However, we find that the dwarf SEL percentage from the NSA catalog (using our new scheme) is the highest despite having MS11$\rm_{solar}$-based fluxes, possibly due to the additional flux calibrations in the NSA catalog. We would like to stress that all three catalogs are state-of-the-art and that we have no evidence that the differences in statistics are due to mistakes in any of the catalogs. Rather, the large discrepancies in emission-line-based AGN statistics are an important result of this paper, associated with the many methodological choices made during the spectral fitting process. Such discrepancies are unavoidable without consensus on methodological choices and should be represented in any statistical conclusions. \\

In this paper, we display the MPA-JHU SEL sample in all plots since this sample has the most galaxies and its SEL properties lie between the properties from the other two catalogs. However, we cannot determine whether any catalog is clearly better, so we report the statistics of all three catalogs.   \\ 

\section{Physical Properties of AGN Candidates} \label{sec:physical}
We explore the physical properties of the AGN host galaxies identified by our optimized scheme with a focus on dwarf AGN. 
\subsection{Gas Content and Metallicity}
\begin{figure}
    \centering
    \includegraphics[width=\linewidth]{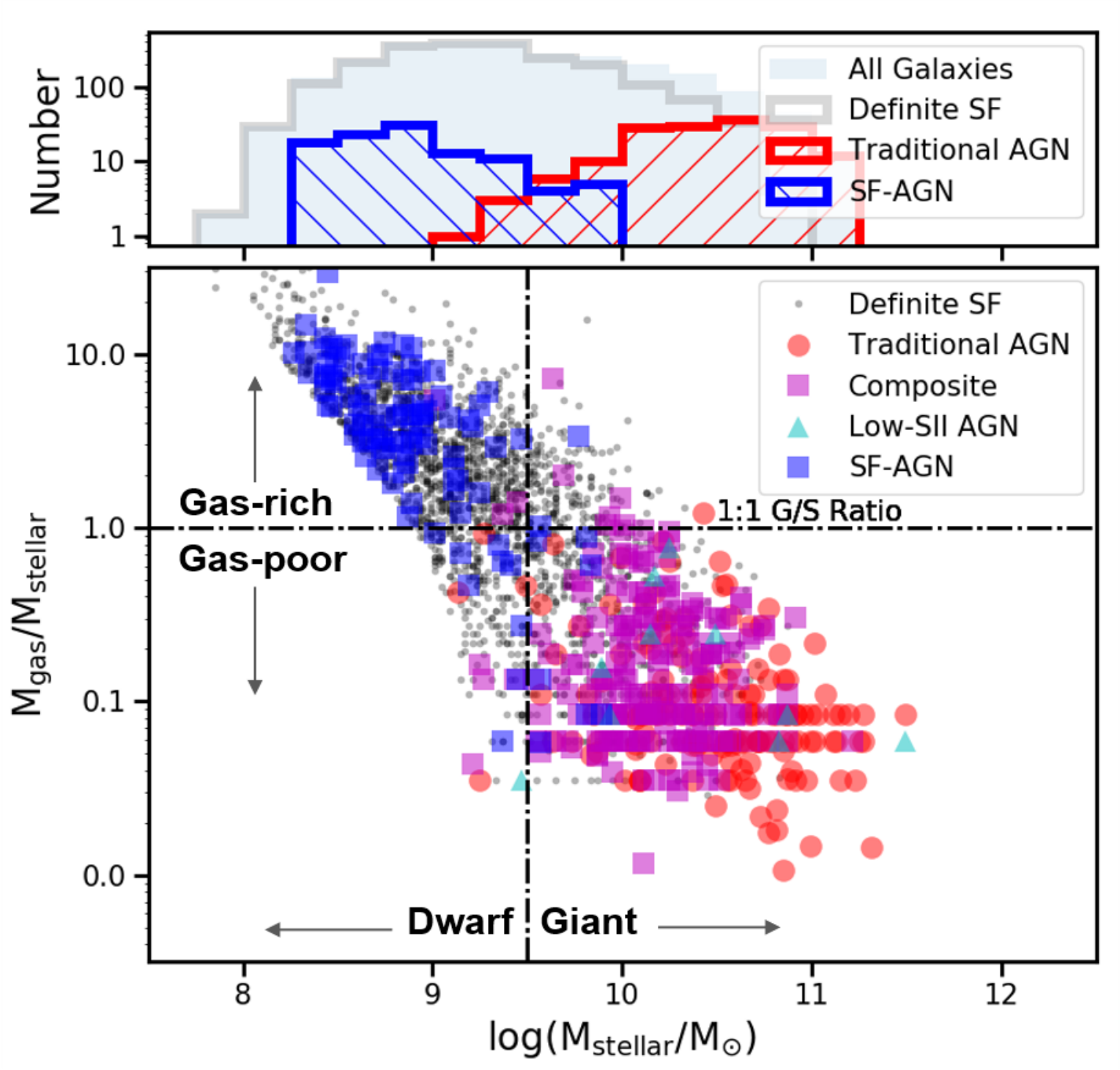}
    \caption{Gas content of RESOLVE and ECO SEL galaxies. The horizontal line represents a gas-to-stellar-mass ratio of 1. The vertical line is the gas richness threshold mass (\mstar $\sim 10^{9.5}$ \msun), which is our definition of the dwarf-giant divide. Almost all SF-AGN (blue squares) are gas-dominated SEL dwarfs, while Composites (magenta squares) and Traditional AGN (red circles) are mostly gas-poor giants.}
    \label{fig:physical_gas}
\end{figure}
\begin{figure}
    \centering
    \includegraphics[width=\linewidth]{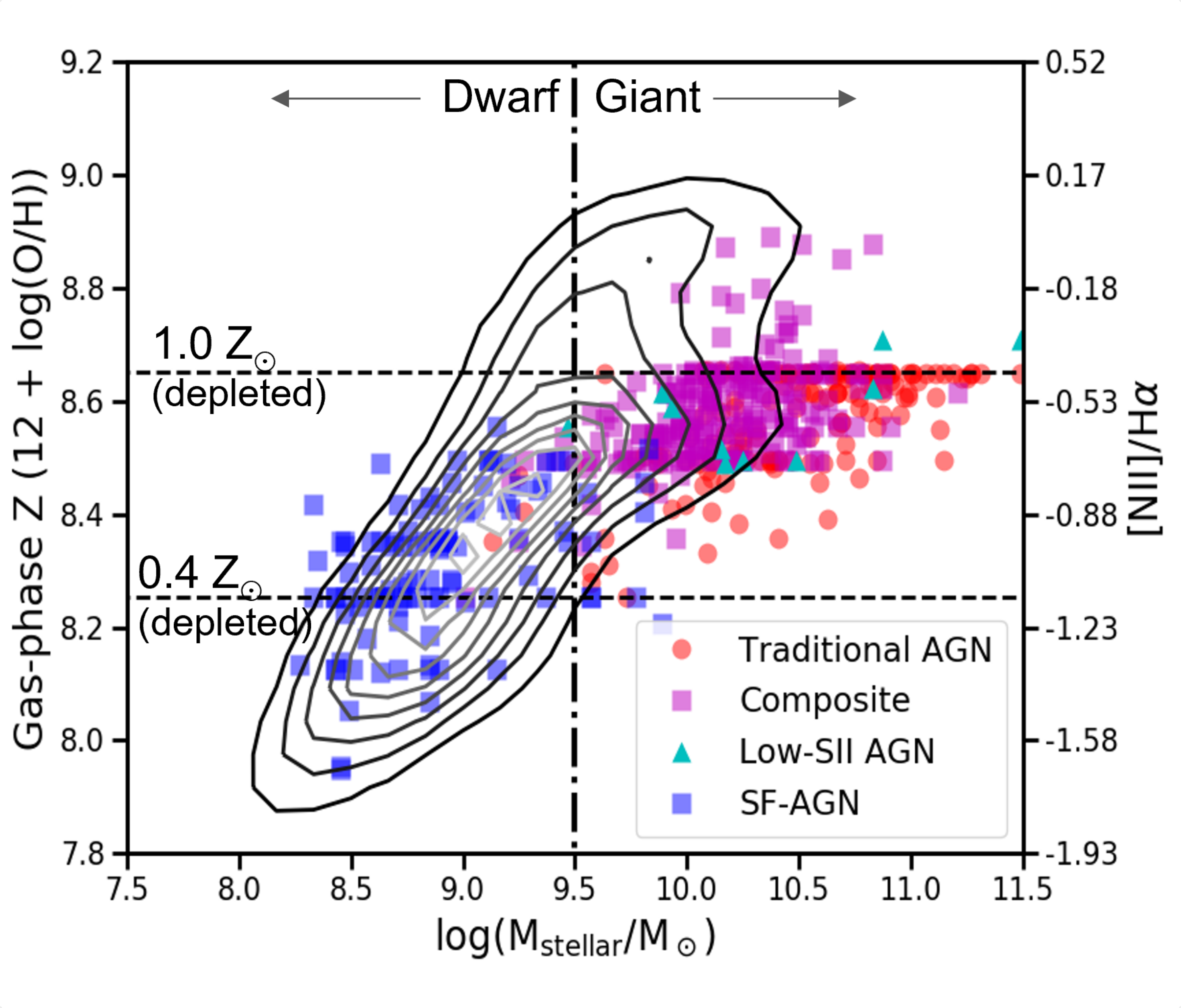}
    \caption{Mass-metallicity relation of RESOLVE and ECO SEL galaxies, showing that almost all SF-AGN (blue squares) are low metallicity SEL dwarfs with a median metallicity of 0.45 \zsun. Metallicities are estimated using NebulaBayes and our Cloudy photoionization grids using BPASS stellar continuum models. While inputting to Cloudy, we convert the stellar metallicity-based continuum models to gas-phase metallicities in 12+log[O/H] units by applying a depletion factor of 0.11 dex. The horizontal lines represent metallicities of 0.4 \zsun and 1 \zsun with our chosen depletion factor of 0.11 dex applied (1 \zsun (12 + log[O/H]) = 8.76 - 0.11 = 8.65). For reference, the right vertical axis provides the equivalent \NvH corresponding to the left axis metallicity using the relation from \citet{Pettini04}. Note that we follow the general literature practice of applying the [NII]/H$\alpha$ calibration for SF galaxies (majority of the sample) to all galaxies in order to plot all the galaxies together. The vertical line is as in Figure \ref{fig:physical_gas}. The contours represent the number density of Definite SF galaxies at the 10th, 20th, ... 90th percentile levels.}
    \label{fig:met}
\end{figure}
Figure \ref{fig:physical_gas} shows the gas-to-stellar mass ratios (G/S) of all RESOLVE and ECO SEL galaxies. Most of the Traditional AGN and Composites are in gas-poor giant galaxies, and almost all SF-AGN are in gas-dominated dwarfs. The high G/S of SF-AGN is typical of dwarfs in the local Universe \citep{Kannappan04, Kannappan13, Stark16}. \\

Figure \ref{fig:met} shows the mass-metallicity relation of RESOLVE and ECO SEL galaxies, using the gas phase metallicities obtained from the Bayesian inference code, NebulaBayes, as detailed in Section \ref{subsec:metallicity}. The median metallicity of Composites and Traditional AGN, $0.8$ \zsun, is higher than the median metallicity of the dwarf-dominated RESOLVE SEL sample as a whole, 0.7 \zsun. On the other hand, most SF-AGN are hosted by dwarfs and have a median metallicity of $\sim$0.45 \zsun, slightly higher than the fiducial dwarf metallicity used in our models in Section \ref{sec:models}. We find that SF and AGN galaxies follow different mass-metallicity trend lines, as has been observed in other work \citep[e.g.,][]{Thomas19}. However, the metallicities of  AGN  in this work are not optimally modelled because we use a pure SF photoionization grid. We recognize that the metallicities would change if our modelling included AGN contributions. However, adaptive modelling of AGN metallicities  is beyond the scope of this paper, and Figure \ref{fig:met} is purely demonstrative of trends.\\

Importantly, despite SF-AGN being selected \textit{only} based on differential classifications between the diagnostic diagrams, between 75-95\% of SF-AGN (depending on catalog) belong to the category of low-metallicity gas-rich dwarfs.  \\

\subsection{Star Formation}\label{subsec:sfr}
\begin{figure}
    \centering
    \includegraphics[width=\linewidth]{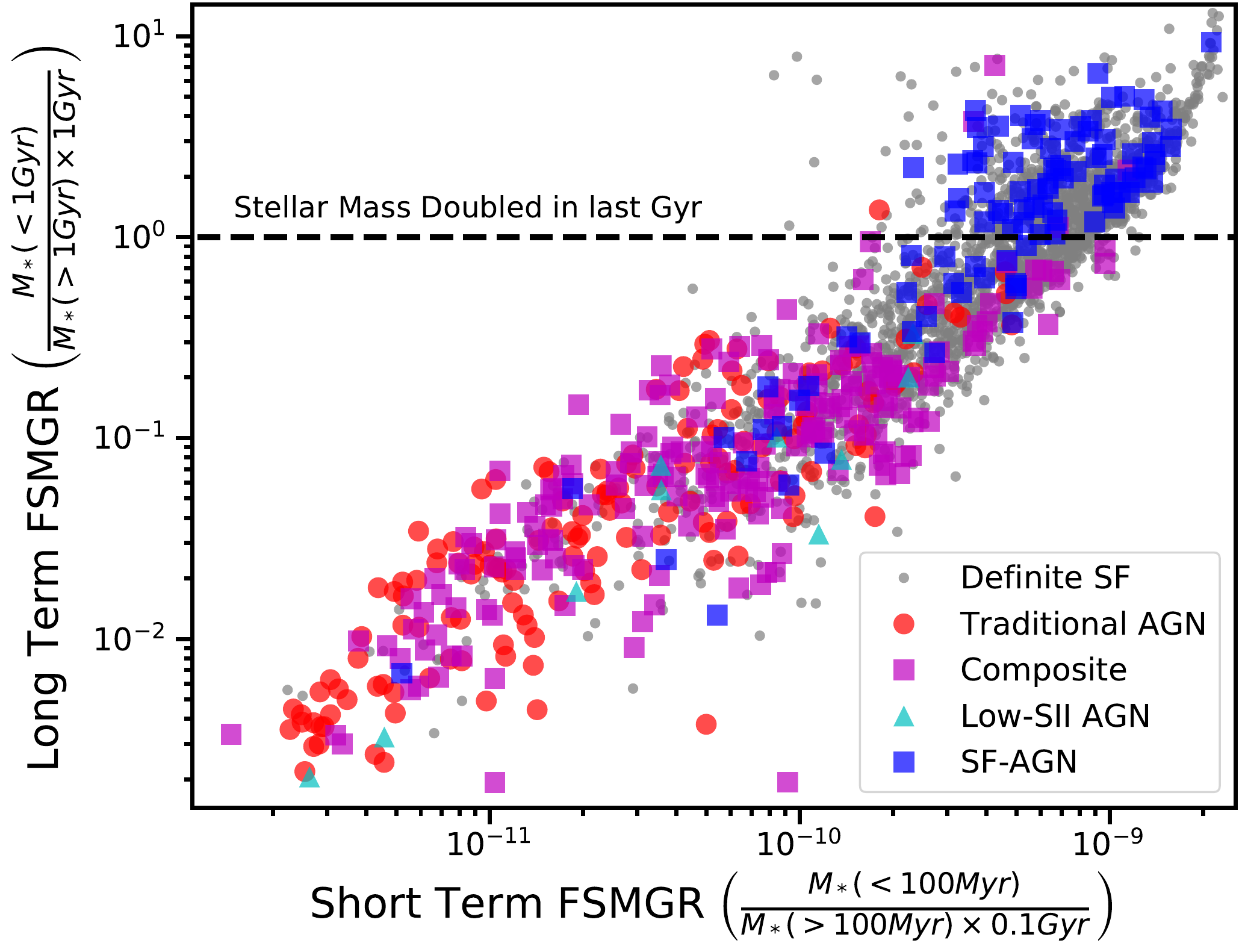}
    \caption{Long Term (1 Gyr) vs. Short Term (100 Myr) star formation histories of RESOLVE and ECO SEL galaxies as measured using FSMGR (see Section \ref{subsec:surveydata}). The horizontal line is where galaxies are doubling their stellar mass in a Gyr. Most SF-AGN have much higher star formation activity than Composites or Traditional AGN.}
    \label{fig:ssfr}
\end{figure}
Figure \ref{fig:ssfr} shows that in our SEL sample, Composites and Traditional AGN have low long- and short-term star formation rates. We trace star formation histories using long- and short-term fractional stellar mass growth rates (FSMGR), defined as the ratio of newly formed stellar mass to preexisting stellar mass per timescale, where the timescale dictating the division of new and preexisting is at 1 Gyr and 100 Myr, respectively (see Section \ref{subsec:surveydata}). \\

Most SF-AGN are gas-rich dwarfs that have more than doubled their stellar mass in the past Gyr lying above the line in Figure \ref{fig:ssfr}, so they are much more actively star forming than Composites and Traditional AGN. This high star formation activity implies a dilution of AGN signatures and consequently makes them hard to identify with the NII and SII plots, as discussed by \citet{Moran02}, \citet{Reines13}, and in Section \ref{subsec:need} and Figure \ref{fig:gridsimple}. \\

We note that the handful of outliers in the SEL sample at low FSMGR$\rm_{LT}$ and moderate FSMGR$\rm_{ST}$ are part of a larger population of very dusty galaxies whose FSMGR$\rm_{ST}$ is well measured from UV and IR photometry, but whose FSMGR$\rm_{LT}$ is likely underestimated by SED fitting that does not include mid-IR photometry representing dusty SF (see Section \ref{subsec:surveydata}). These outliers are not associated with mid-IR detected AGN (see Paper II). \\

\subsection{Group Halo Properties}
\label{subsec:grphaloprop}
\begin{figure*}
    \centering
    \includegraphics[width=\linewidth]{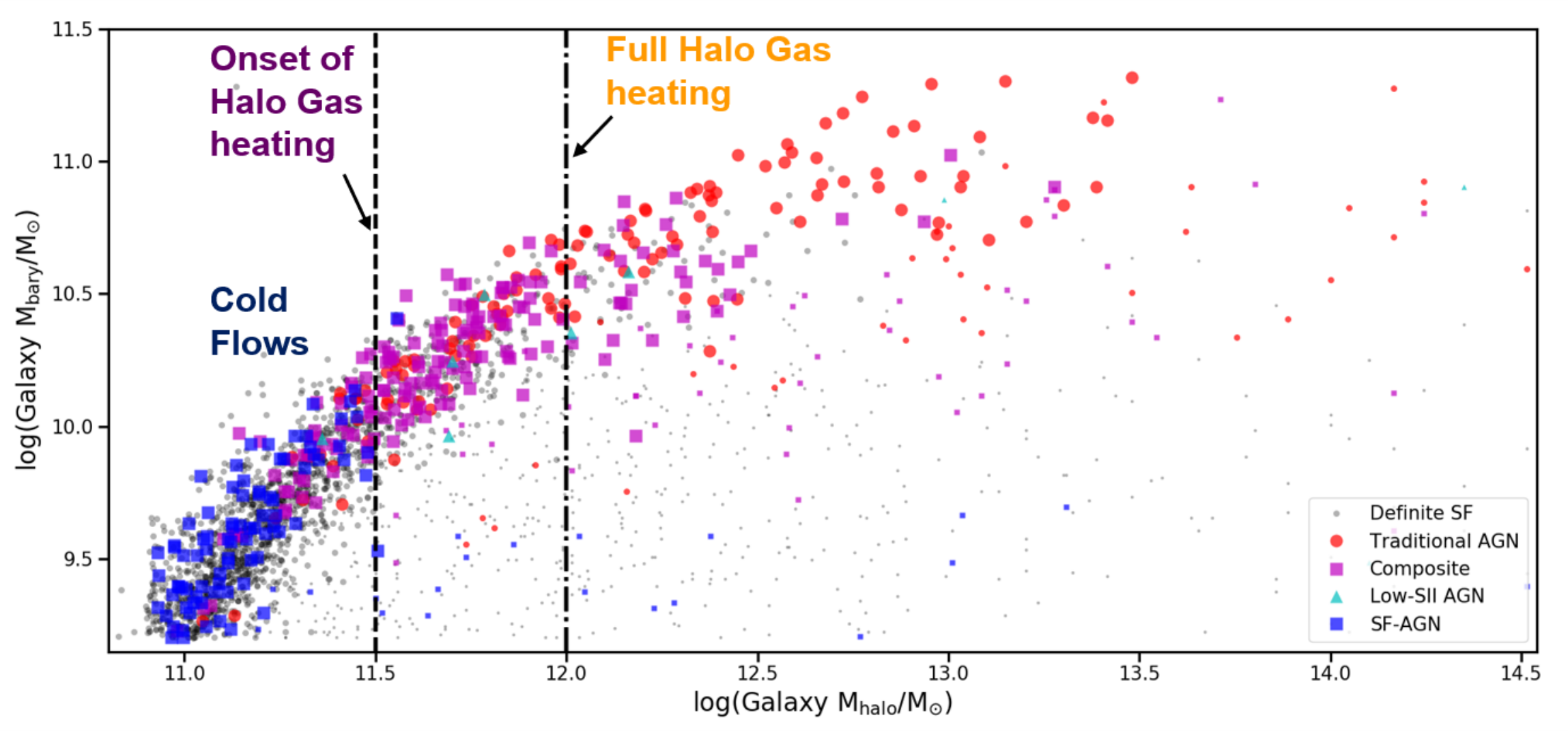}
    \caption{Galaxy baryonic mass vs. group halo mass properties of RESOLVE and ECO SEL galaxies. The dashed line at M$\rm_{halo} \sim 10^{11.5}$\msun represents the gas-richness threshold mass scale and the dot-dashed line at M$\rm_{halo} \sim 10^{12.0}$\msun represents the bimodality mass scale, as discussed in Section~\ref{subsec:grphaloprop}. The larger and smaller symbols represent central and satellite galaxies respectively. There is an evident transition in AGN categories at key mass scales: SF-AGN below the gas-richness threshold mass scale, Traditional AGN above the bimodality scale, and Composites in the transition zone. Most SF-AGN are hosted by dwarfs in single-galaxy halos, in contrast to Traditional AGN hosted by more massive galaxies in multi-galaxy halos.}
    \label{fig:grouphalo}
\end{figure*}
Figure \ref{fig:grouphalo} investigates the relationship between the group halo mass and galaxy baryonic mass of RESOLVE and ECO SEL galaxies. We see shifts at two important mass scales -- the gas-richness threshold mass, defined as M$\rm_{halo} \sim 10^{11.5}$ \msun, which corresponds to \mbary $\sim 10^{9.9}$ \msun and \mstar $\sim 10^{9.5}$ \msun \citep{Dekel86, Kannappan13}, and the bimodality mass, defined as M$\rm_{halo} \sim 10^{12}$ \msun \citep{Dekel06}, which corresponds to \mbary $\sim 10^{10.6}$ \msun and \mstar $\sim 10^{10.5}$ \msun \citep{Kauffmann03mass, Kannappan13}. The gas-richness threshold scale marks the onset of gas heating within 0.1 $R_{virial}$ of the dark matter halo, while the bimodality scale corresponds to the mass where the entire halo is shock heated \citep{Dekel06}. Figure \ref{fig:grouphalo} shows a sharp transition from AGN candidates being mostly SF-AGN below the gas richness threshold mass to their being mostly Traditional AGN above the bimodality mass. Composites span the transition range and overlap the other AGN categories. We also find that SF-AGN mainly occupy single-galaxy low-mass halos, whereas Traditional AGN are more commonly found in multi-galaxy massive halos.\\

\section{Discussion} \label{sec:discussion}
Table \ref{tab:litreview} provides a summary of statistics from a number of previous dwarf AGN studies along with our own work. There has not been a clear consensus regarding the \zzero dwarf AGN frequency among various systematic searches. Recent studies of \zzero dwarfs using mid-IR colors, have estimated the dwarf AGN percentage to be between 0.2\% to 20\% \citep{Sartori15, Hainline16, Kaviraj19, Lupi20}. In the optical, by using narrow emission lines with the BPT diagram and/or by finding broad \ha emission, the dwarf AGN percentage has been estimated to be between 1\% to 3\% \citep{Reines13, Sartori15, Bradford18}. Multiple X-ray studies have used the $\rm L_{X-ray}-SFR$ relationship to estimate local dwarf AGN percentages to be between 0.1\% to 3\% \citep{Schramm13,Lemons15,Pardo16, Birchall20}. In a radio study, \citet{Reines19} used the $\rm L_{radio}-SFR$ relation to determine that AGN hosts make up  $\sim$12\% of their \zzero dwarf galaxy sample. An important point to note is that each of these studies has different sample selection criteria and thus is representative of different sub-populations of dwarfs.\\

\startlongtable
\begin{longtable*}{ C{1cm} | C{1.2cm} |C{1.6cm} |C{1.75cm} | C{1.75cm} |C{2.2cm} |C{1.6cm}|C{3.0cm}|C{2.0cm}}
\caption{Statistics from previous dwarf AGN studies \label{tab:litreview}}\\
\hline 
\hline
Spectral Range & Name & Data Source & Sample selection & \# of dwarfs & AGN selection & \# of AGN (\% AGN in sample 
dwarfs) & Sample biases & Notes on AGN \\
\hline
\endhead
\multirow{-5}{1cm}{Mid-IR}                                                 & \citet{Hainline16} & SDSS - NSA catalog; AllWISE          & \mstar$<3\times10^{9}$\msun, z \textless 0.055, S/N $>$ 3 in W1, W2 and W3                                                                                      & 18000          & mid-IR color cut \citep{Jarrett11}  & $\sim$41 ($\sim$0.2\%) & mid-IR AGN selection can be contaminated by extreme SF especially if using only W1 - W2 color cut  &  W2 - W3 colour important for selecting dwarf AGN to minimize SF contamination    \\ 

\cline{2-9}     
                    & \citet{Kaviraj19} & HSC; AllWISE          & \mstar$<10^{9}$\msun, 0.1 \textless z \textless 0.3, S/N $>$ 5 in W1 and W2                                                                                      & N/A          & mid-IR color cut \citep{Satyapal14, Satyapal18}  & $\sim$800 (10-30\%) & mid-IR AGN selection can be contaminated by extreme SF or can miss low-\textit{z} obscure AGN;  & Mergers not important for triggering dwarf AGN       \\
\cline{2-9}
\multirow{-10}{1cm}{Mid-IR}                                                 & \citet{Lupi20} & HSC; ALLWISE          & \mstar$<10^{9}$\msun, 0.1 \textless z \textless 0.3, S/N $>$ 5 in W1 and W2, S/N $>$ 2 in W3                                                                                  & around 5000 & mid-IR color cut \citep{Satyapal14}  & 20 (0.4\%) & mid-IR AGN selection can be contaminated by extreme SF or can miss low-\textit{z} obscure AGN, especially with low resolution WISE photometry & Better cross-matching between surveys and higher S/N cut on W3 data yields lower dwarf AGN percentage than in \citet{Kaviraj19}.    \\                                     
\hline                                                
\multirow{-8}{1cm}{Mid-IR and Optical narrow emission lines} & \citet{Sartori15} & SDSS - MPA-JHU catalog; WISE    & \mstar$<10^{9.5}$ \msun; z \textless 0.1   & 48000  & BPT plot, He II plot, mid-IR color cut \citep{Jarrett11, Stern12} & 336: 47 BPT, 112 He II, 77 mid-IR (0.70\%) & BPT plot biased towards finding AGN in high-Z giants; only small sample with strong He II; mid-IR selection can be contaminated by extreme SF & Only 3 AGN identified by all 3 methods\\
\hline
\multirow{5}{1cm}{Optical narrow emission lines} & \citet{Bradford18} & SDSS - NSA; Arecibo           & z \textless 0.055; isolated; HI data measured \& BPT lines S/N $> 3$ & 867          & distance from SF sequence on BPT plot, d$\rm_{BPT} >$ 0.11 dex & 24 (2.80\%)                                 & Incomplete HI flux-limited sample    & Either AGN or stellar feedback quenched HI gas in dwarfs\\
\cline{2-9}
                                                 & This work & RESOLVE and ECO surveys; SDSS - MPA-JHU, Portsmouth, NSA;  & 0.015 \textless z \textless 0.023; SEL S/N $>$ 5 &  RESOLVE: 226/ 129/ 114 \& ECO: 1525/ 749/ 738 in MPA-JHU/ Portsmouth/ NSA & optimized scheme using NII, SII, and OI plots & RESOLVE: 6.6\%/ 3.1\%/ 15.0\% \& ECO: 6.9\%/ 6.0\%/ 15.6\% in MPA-JHU/ Portsmouth/ NSA  & Requiring SELs biases against finding AGN in non-emission line galaxies    & New category called SF-AGN in metal-poor gas-rich hosts accounts for most of the previously undetected dwarf AGN \\
\hline
\multirow{-7}{1cm}{Optical narrow and broad emission line} & \citet{Reines13}    & SDSS - NSA      & \mstar\textless$10^{9.5}$\msun; z\textless{}0.055; S/N $>$ 3 and EW $>$ 1 for \ha, [N II], [O III], S/N $>$ 2 for \hb & 25974        & BPT plot, broad Ha                                                                                       & 151: 136 BPT, 15 broad Ha (0.58\%)          & Sample biased towards luminous galaxies and AGN; low-Z AGN overlaps with SF wing of BPT; Both methods not sensitive to LMBH in blue star forming dwarfs & low-mass broadline AGN fall in SF wing of BPT \\
\hline
\multirow{5}{1cm}{Optical broad emission line}   & \citet{Greene07}  & SDSS DR4              & z \textless 0.352; high rms above continuum near \ha  & N/A          & broad \ha detected using custom spectral fitting pipeline & 174 (N/A)                                   & Selection effects -- biased against finding IMBHs at low accretion rates       hosts                   & AGN hosts typically have recent SF\\
\cline{2-9}
                                                 & \citet{Dong12}    & SDSS DR4              & SDSS ``galaxy" or ``QSO"; z \textless 0.35    & N/A          & broad \ha detected using custom spectral fitting pipeline                                        & 306 (N/A) & Survey selection effects -- biased against finding low accretion IMBHs & Recursive pipeline finds low L$\rm_{X-ray}$ AGN with low accretion rates \\
\hline
\multirow{-3}{1cm}{X-ray}                       & \citet{Schramm13} & GEMS survey - Chandra; HST      & z \textless 0.3; \mstar \textless $10^{9.5}$\msun; X-ray detection & 2100 & X-ray luminosity vs. SFR & 3 (0.1\%) & 
                                                Selection effects -- bias towards higher X-ray luminosity AGN & One AGN candidate has broad \ha \\
\cline{2-9}
                        & \citet{Lemons15}  & SDSS - NSA; Chandra & z \textless 0.055; \mstar \textless $10^{9.5}$\msun; Chandra crossmatch  & 44594 & L$\rm_{X-ray}$ \textgreater expected L$\rm_{XRB}$                                         & 19 (0.05\%)                                 & Biased towards more massive accreting BHs (Chandra's sensitivity) & X-ray AGN hosts are physically small (r$\rm_{50}$ \textless 2kpc)\\
\cline{2-9}
                                                 & \citet{Pardo16}   & NEWFIRM; DEEP2; Chandra         & $0.1<$z$<0.6$; \mstar\textless{}$10^{9.5}$\msun & 605          & 
                                                 L$\rm_{X-ray}$ vs. SFR; L$\rm_{X-ray}$ \textgreater expected L$\rm_{XRB}$ & 10 (0.6-3\%) & Incomplete below L$\rm_{X-ray} < \sim10^{41} erg s^{-1}$; & AGN fraction agrees with SAM prediction\\
\cline{2-9}
                                                 & \citet{Latimer19}  & FIRST, VLA, Chandra, SDSS - NSA & BCDs from \citet{GildePaz03} with FIRST detections and d $<$ 20 Mpc OR detectable hard X-ray point sources from Chandra	& 5	& X-ray Luminosity vs. SFR;  radio luminosity of compact source	&  1 (20\%) & Small sample size & One BCD could host candidate low-luminosity AGN if the spatially coincident X-ray and radio emissions are coming from same source \\
\cline{2-9}
                          & \cite{Birchall20} & SDSS -- MPA-JHU; 3XMM DR7 & MPA-JHU overlapping with 3XMM; z$<$0.25; $M_* < 3\times10^9 M_\odot$ & 4331 & $L_{X-ray} > 3(L_{XRB} + L_{X-ray gas})$ & 61 (1.4\%)
                          & Sample biased toward luminous AGN with higher accretion rates & $\sim$85\% of X-ray AGN are not classified as AGN by the NII plot. \\
\hline
Radio                                            & \citet{Reines19}   & SDSS - NSA ; VLA; FIRST         & z \textless 0.055; \mstar \textless $10^{9.5}$\msun; FIRST crossmatch & 111 & L$\rm_{9GHz}$ vs. expected L$\rm_{9GHz-SNe}$ & 13 (11.7\%)                                                                 & Many AGN do not produce radio continuum emission detectable by FIRST & Most AGN offset from center, and SF in BPT plot, but AGN in OI plot \\ 
\end{longtable*}

Our dwarf AGN frequency of 3-16\% falls on the higher end of most dwarf AGN frequency ranges. At first glance, the high mid-IR dwarf AGN percentage ($\sim$20\%) from \citet{Kaviraj19} seems to be close to our dwarf AGN percentage from the NSA catalog ($\sim$16\%). However, \citet{Lupi20} have found that the mid-IR AGN percentage drops to $\sim$0.4\% if stricter cross-matching and S/N criteria are imposed on the WISE photometric data. This low percentage is in agreement with results from both \citet{Sartori15} and \citet{Hainline16} who find that mid-IR color selection for AGN suffers from high levels of contamination from strong SF, especially when using only the W1-W2 colors. In Paper II, we will compare optical and mid-IR-selected dwarf AGN in the RESOLVE and ECO surveys. \\

The dwarf AGN percentage of $\sim$12\% from the radio study by \citet{Reines19} is close to the upper end of our optical dwarf AGN percentage range. Their dwarf sample is selected based on VLA FIRST radio continuum detections that overlap with dwarfs in the SDSS NSA catalog. However, the authors note that radio continuum detections at FIRST sensitivity levels are rare for local dwarfs; only $\sim$0.3\% of their parent sample of 43,707 dwarfs have FIRST detections. Based on their data, we calculate that the radio dwarf AGN percentage normalized to the full dwarf population in their parent survey is only $\sim$0.03\%. \cite{Birchall22} find a higher completeness-corrected full-population
dwarf AGN percentage of $\sim$1\% in subsequent
analysis of X-ray AGN identified in \cite{Birchall20}. In comparison, the optical SEL dwarf AGN percentage normalized to the full dwarf population in RESOLVE and ECO is $\sim$0.6-3.0\% based on our new optimized scheme (see Section \ref{subsec:agnstats}). We note that the RESOLVE/ECO full-population numbers are normalized to a volume- and mass-limited full-population, whereas the other ``full-population" numbers represent flux-limited subsample AGN frequencies normalized to flux-limited parent samples. \\

The searches that are most similar to this work in methodology are previous studies that use optical emission line diagnostics. Using the NII plot alone, \citet{Reines13} find $\sim$0.5\% of dwarfs in their emission line sample have AGN signatures (Composites, Seyferts, and LINERs). Their filtering criteria are more relaxed than ours; they use S/N $\geq$ 2 for \hb and S/N $\geq$ 3 for [N~II], [O~III], and \ha, while we use S/N $>$ 5 for all aforementioned lines along with [S~II] and [O~I]. We find that decreasing the S/N threshold from 5 to 3 for the emission lines in the NII plot increases the number of galaxies in our sample by only $\sim$15\%. The main difference between the samples in this work and in \citet{Reines13} is the strict cut that we impose on [O~I] fluxes, which makes our sample much smaller and more biased towards bluer colors. Yet our method finds more AGN overall by using the NII, SII, and OI plots together. Normalizing the statistics from \citet{Reines13} to the full dwarf population they select from the SDSS parent survey (44,594 dwarfs), their narrow-line dwarf AGN percentage is $\sim$0.3\% compared to $\sim$0.6-3.0\% in this work. Similarly, \citet{Sartori15} report that among the $\sim$48,000 dwarfs in their parent sample, $\sim$0.1\% are AGN hosts using the NII plot alone; adding the HeII diagnostic yields an AGN percentage of $\sim$0.3\%. Despite their using similar selection criteria to \citet{Reines13}, they report a lower AGN percentage using the NII plot alone since they consider only Seyferts to be AGN candidates. \citet{Sartori15} find twice as many AGN using the HeII diagnostic than just the NII plot alone, but their parent-sample normalized percentage of optical dwarf AGN ($\sim$0.3\%) is still lower than our full-population dwarf AGN percentage ($\sim$0.6-3.0\%) using the NII, SII, and OI plots together. Among all the studies we compare to, the only studies that explicitly use the \OvH line ratio from SDSS data are those by \citet{Kewley2006} and \citet{Reines19}. We cannot directly compare dwarf AGN percentages with \citet{Kewley2006} since they do not report dwarf statistics, but their stellar mass histograms show virtually no dwarfs in any of their non-SF categories (which exclude SF-AGN, see Section~\ref{subsec:need}). \citet{Reines19} examine the optical emission line classifications of compact radio AGN and find that 1 out of 13 compact radio AGN is classified as an optical AGN by all three diagnostic plots, NII, SII, and OI. Interestingly 5 out of 13 compact radio AGN are classified as optical AGN by the OI plot, but not by the NII plot, so we would label these SF-AGN. Overall, our dwarf AGN percentage is higher than most previous estimates, especially from optical studies. \\

We also compare our dwarf AGN percentage with AGN populations in various simulations described by \citet{Haidar22}. The overall RESOLVE and ECO dwarf AGN percentage of $\sim$0.6-3.0\% (for the full mass and volume-limited surveys, not just for SELs) agrees with the dwarf AGN occupation fraction seen in the Horizon-AGN simulation in all X-ray luminosity bins, and with EAGLE, Illustris, and TNG100 in some X-ray bins (Figure 5 of \citealt{Haidar22}). However, \citet{Haidar22} highlight several sources of overestimation or underestimation of AGN percentages for both theory and observations. On the theoretical side, simulations generally neglect BH wandering and obscuration, thus \textit{overestimating} accretion and its observability, and seeding mechanisms are unphysical. On the observational side, optical emission line methods can miss AGN that are off-center, obscured, and/or have low luminosities, \textit{underestimating} the dwarf AGN percentage. On the other hand, past SEL studies have generally reported the AGN frequency for SEL galaxies, which \textit{overestimates} the SEL-detected AGN frequency of the full galaxy population. \citealt{Haidar22} stress the need for full-population statistics for a fair comparison between simulations and observations. Our raw SEL AGN frequencies exceed the frequencies to which \citet{Haidar22} compare (e.g., from \citealt{Reines13}). Our full-population normalized SEL AGN frequencies, which are by definition lower than our raw SEL AGN frequencies and more directly comparable to simulations, also exceed previous estimates (where possible, as for \citealt{Reines13} and \citealt{Sartori15} in Table \ref{tab:litreview}). Detection methods sensitive to obscured, wandering, and low-luminosity AGN represent the only robust source of higher AGN frequencies. In Paper II, we will use two other AGN identification methods suited for dwarfs, yielding a more comprehensive census of local dwarf AGN. \\

We note that even a complete dwarf AGN census cannot yet help constrain BH seed formation mechanisms. First, we are working at \zzero where even dwarf BHs may have evolved significantly beyond seeds. Second, theoretical simulations do not yet include realistic BH seeding or evolution \citep{Haidar22}. Third, observationally feasible AGN detection methods applied to the full galaxy population have not yet found AGN percentages that approach the relevant BH percentages to differentiate between the two possible mechanisms. The two leading theoretical mechanisms for forming seed BHs predict occupation fractions of 100\% (for `light' seeds) vs. 85\% (for `heavy' seeds) in dwarfs \citep{Volonteri08, Greene12, Natarajan14}, more than two orders of magnitude higher than the AGN occupation fraction in the entire RESOLVE and ECO surveys. Finding more dwarf AGN is key to being able to compare observed \zzero AGN to simulated BHs and to understand how they evolve and grow with their hosts.\\

\section{Conclusions and Future Work}
\label{sec:conclusion}
We have created an optimized galaxy classification scheme that robustly classifies all galaxies into unique categories using a combination of existing optical emission line diagnostic plots. This scheme allows for the identification of AGN in metal-poor dwarfs using the metallicity-insensitive [O~I] and [S~II] diagnostic plots in tandem with the metallicity-sensitive \niiplot. In order to apply our classification scheme to the volume- and mass-limited RESOLVE and ECO surveys, we use emission line fluxes from three SDSS catalogs (MPA-JHU, Portsmouth, and NSA). Applying S/N cuts to these catalogs, we select subsamples of strong emission line (SEL) galaxies, which comprise $\sim$16-35\% of cross-matched galaxies in the combined RESOLVE and ECO surveys limited to \mbary $> 10^{9.2}$ \msun.  
\begin{itemize}
    \item[1.] Our photoionization modelling shows that the OI plot is better than the commonly used NII plot (a.k.a. the BPT diagram) at identifying AGN in galaxies that are metal-poor and/or extremely star forming (Figures \ref{fig:csfandsspgrid} and \ref{fig:gridsimple}). 
    
    \item[2.] Our newly optimized classification scheme (Figure \ref{fig:classification}) classifies galaxies into the following mutually exclusive categories: Definite SF, Composite, Seyfert, LINER, Ambiguous-type AGN, SF-AGN, Low-SII/Low-OI/Low-SII+OI AGN. 
    
    \item[3.] Due to the systematic classification of all galaxies, the optimized scheme identifies a new category of AGN called SF-AGN in metal-poor, gas-rich, star-forming dwarfs (Figures \ref{fig:physical_gas}, \ref{fig:met}, \ref{fig:ssfr}). SF-AGN are missed by traditional AGN identification methods.

    \item[4.] SF-AGN are mostly found in single-galaxy halos in the rapid halo gas cooling regime with \mhalo $< 10^{11.5}$ \msun (the gas-richness threshold mass) whereas Traditional AGN are mostly found in larger groups in the hot halo gas regime with \mhalo $> 10^{12}$ \msun (the bimodality mass; Figure \ref{fig:grouphalo}).
    
    \item[5.] We conclude that SF-AGN are true AGN candidates as their properties and trends in emission line ratios cannot be explained by other phenomena like SF, shocks, or DIG. Additionally, high S/N IFU data from SAMI for one SF-AGN show AGN-like \OvH ratios only in the central 2 kpc. 
    
    \item[6.] Considering SF-AGN, Composites, Seyferts, LINERs, Ambiguous-type AGN, and Low-SII AGN as candidate AGN, we report that the overall AGN percentage in \zzero SEL galaxies is $\sim$16-30\% depending on the SDSS catalog used (Table \ref{table:fullstats}). 
    
    \item[7.] The new frequency of AGN in \zzero SEL dwarfs (i.e., the search sample for this study) is $\sim$3-16\% (Table \ref{table:fullstats} and Figure \ref{fig:dwarfpercent}) versus $<$1\% using traditional optical AGN identification methods. The frequency of dwarf SEL AGN strongly depends on the processing methodology of the data set used, especially differences in spectral decomposition, flux calibration, and associated error analysis. Regardless, our SEL dwarf AGN percentage is on the high end of previously reported dwarf AGN  percentages.
    
    \item[8.] Normalized to the full galaxy population of RESOLVE and ECO (including non-SEL galaxies), $\sim$0.6-3.0\% of dwarfs are AGN hosts as per our optimized classification scheme. This percentage is also much higher than most previous estimates, especially in the optical. 
\end{itemize}

The RESOLVE and ECO SEL surveys are volume-limited and 97\% complete above our baryonic mass floor, hence the dwarf AGN frequencies above are representative of the true population of SEL galaxies that is minimally mass or luminosity biased.    The unexpectedly high frequency of AGN we find in otherwise ordinary dwarf galaxies has potential implications for the mechanisms and evolutionary importance of feedback in dwarfs. \\

Recent theoretical work suggests that some dwarf AGN can be fundamentally different from AGN in giants. These dwarf AGN may not have feedback that is powerful enough to limit star formation, but may instead be limited in gas accretion due to strong stellar feedback \citep{Habouzit17, Angles-Alcazar17, Trebitsch18}. In this star-formation-limiting-AGN scenario, dwarf AGN may be intermittently fuelled as strong star formation feedback -- typical of gas-rich, low-metallicity dwarfs -- fluctuates. These AGN may be too weak to drive ionized gas outflows resulting in broad H$\alpha$, as seen in dwarf AGN found by \citet{Bradford18} and \citet{Latimer19}. Since we have no current evidence of \ha outflows in SF-AGN, our SF-AGN could in principle follow the theoretical star-formation-limiting-AGN-scenario. However, SF-AGN do not have lower short-term SFRs (traced by FSMGR$\rm_{ST}$) than expected from their long-term SFRs (traced by FSMGR$\rm_{LT}$; Figure \ref{fig:ssfr}). Thus, we do not yet have evidence for a recent abatement of SF with the data that we currently have. Nonetheless, further investigation of stellar vs. AGN outflow signatures can test whether feedback in dwarfs and giants are indeed different.  \\

This work has proven the advantages of using multiple emission lines to detect dwarf AGN, as we report a much higher full-population normalized percentage of dwarf AGN than previously seen in optical studies. In Paper II, we will provide detailed comparisons of our optimized scheme to other dwarf AGN detection methods like mid-IR color selection and the redefined BPT demarcation line of \citet{Stasinska06}. \\

\section*{Acknowledgements}
We would like to thank Romeel Dave for his help pointing us to relevant results from simulations, Christopher Agostino for his help with handling data from XMM-Newton, and Ingyin Zaw for her insight on the differences between SDSS catalogs. We are grateful to Michael Palumbo III, Trystyn Berg, and the anonymous referee for providing valuable feedback on the manuscript.  We thank Derrick Carr, Zack Hutchens, and Ella Castelloe for useful discussions at various stages of the project. We acknowledge the efforts of Margie Bruff in the initial phase of this project. This research has been supported by the National Science Foundation under award AST-2007351. MSP acknowledges additional support from the Andrew and Kathrine McMillan Summer Research Fellowship from the UNC Graduate School and the Hamilton Award from the UNC Department of Physics and Astronomy. CR acknowledges the support of the Elon University FR\&D committee and the Extreme Science and Engineering Discovery Environment (XSEDE), which is supported by NSF grant number ACI-1548561, through allocation TG-AST140040. JMB acknowledges support from NSF AST-1812642 and the CUNY JFRASE award. 
\bibliography{References} 

\end{document}